\documentclass[prd,amsfonts,onecolumn,superscriptaddress,aps,nofootinbib,11pt]{revtex4-1}

\usepackage{amsmath, amsthm, amssymb, amsfonts, mathrsfs}
\usepackage{scalefnt}
\usepackage{physics, braket, slashed, bm}
\usepackage{bbold}
\usepackage{listings}
\usepackage{siunitx}
\usepackage{booktabs}
\usepackage{soul}
\usepackage{color}
\usepackage[usenames,dvipsnames]{xcolor}
\usepackage{graphicx}
\usepackage{float}
\usepackage{placeins}
\usepackage{hyperref}
\usepackage[compat=1.1.0]{tikz-feynman}
\usepackage[caption=false]{subfig}
\usepackage{soul}
\usepackage{ulem}

\hypersetup{
	unicode=false,          
	pdftoolbar=true,        
	pdfmenubar=true,        
	pdffitwindow=false,     
	pdfauthor={William},     
	colorlinks=true,       
	linkcolor=blue,          
	citecolor=magenta,        
	urlcolor=blue
}

\begin{document}
	
\title{ 
Same-Sign Taus Signatures of Maximally Flavor-Violating Scalars at the LHC }

\author{Alexandre Alves}
\email{aalves.unifesp@gmail.com}
\affiliation{Departamento de F\'isica, Universidade Federal de S\~ao Paulo, UNIFESP, 09972-270, Diadema-SP, Brazil}
\author{Alex G. Dias}
\email{alex.dias@ufabc.edu.br}
\affiliation{Centro de Ci\^encias Naturais e Humanas, Universidade Federal do ABC,\\
UFABC, 09210-580, Santo Andr\'e-SP, Brazil}
\author{Eduardo da Silva Almeida}
\email{almeidae@ufba.br}
\affiliation{Departamento de Física do Estado Sólido, Universidade Federal da Bahia,\\
UFBA, 40170-115, Salvador-BA, Brazil }
\author{Diego S. V. Gon\c{c}alves}
\email{diego.vieira@ufabc.edu.br}
\affiliation{Centro de Ci\^encias Naturais e Humanas, Universidade Federal do ABC,\\
UFABC, 09210-580, Santo Andr\'e-SP, Brazil}

\date{\today}
	
\begin{abstract}
We explore single and double flavor-violating scalar (flavon) production at the 13 and 14 TeV LHC in an effective field theory formulation where flavons always change the flavor of the Standard Model fermions. When those scalars couple to mass, their flavor-changing couplings to top quarks and tau leptons are favored. Focusing on the mass region below the top-quark mass, we find couplings that fit the muon $(g-2)$ discrepancy and avoid several current experimental constraints. We determine the potential of the LHC to exclude or discover such a new physics scenario with clean signatures consisting of same-sign tau leptons and the simultaneous observation of resonances in the tau plus electron or muon invariant mass. We found that in the double production mode, effective couplings down to order $10^{-2}$ TeV$^{-1}$ can be probed for flavon masses in the 10--170 GeV range at the 14 TeV HL-LHC, but couplings down to 0.1 TeV$^{-1}$ can already be excluded at 95\% confidence level with data collected from the 13 TeV LHC in the same mass interval. We also explore the impact of sizeable diagonal flavon couplings on the prospects of LHC for the signals we propose.
\end{abstract}

\maketitle
	
\section{Introduction}

\setcounter{page}{1}

Spin-zero particles with flavor-violating (FV) interactions are predicted in several models motivated by unsolved questions in the Standard Model (SM). Examples of such particles include pseudoscalars in the form of Nambu-Goldstone bosons, like familons, which are associated with a symmetry among fermion families~\cite{Davidson:1981zd,Wilczek:1982rv,Reiss:1982sq,Gelmini:1982zz,Davidson:1983fy,Feng:1997tn}; majorons, which arise from the breaking of lepton number symmetry in connection to neutrino masses~\cite{Mohapatra:1979ia,Chikashige:1980qk,Chikashige:1980ui,Gelmini:1980re}; axions, which are the pseudo Nambu-Goldstone bosons of an anomalous global chiral symmetry proposed to solve the strong CP problem \cite{Peccei:1977hh,Weinberg:1977ma,Wilczek:1977pj,Georgi:1986df}, and can also be relevant in astrophysics, cosmology, and flavor physics~\cite{DiLuzio:2020wdo,Ema:2016ops,Bjorkeroth:2018dzu}; and axionlike particles (ALPs), which bear some similarities with axions~\cite{Jaeckel:2010ni,Ringwald:2012cu,Dias:2014osa,Bauer:2021mvw,Jho:2022snj}. It is still possible that these particles are linked to the flavor $U(1)_F$ symmetry based on the Froggatt-Nielsen mechanism~\cite{Froggatt:1978nt}, which furnishes an explanation to the fermion hierarchical flavor structure through effectively suppressed Yukawa couplings. In this case, axions and ALPs can be also be identified as flavons~\cite{Ema:2016ops,Bjorkeroth:2018dzu}---particles resulting from the breaking of flavor symmetry. 

As a general feature, the fields associated with these pseudoscalar particles have derivative couplings to the matter fields due to the presence of the shift symmetry $a(x)\,\rightarrow\,a(x)+\mathrm{constant}$ for the pseudoscalar field $a(x)$, and interactions suppressed by the energy scale in which their related global symmetry is broken. For the case in which these particles are very light and with feeble interactions, there are several experiments looking for them, including  astrophysical studies,  focusing on their diagonal couplings with leptons and photons~\cite{Workman:2022ynf,DiLuzio:2020wdo}. In these cases, they cannot be probed directly by colliders. However, it may well be possible that for certain types of ALPs and flavons the new physics energy scale suppressing the interactions is low enough allowing one to test the production of those particles through the proton-proton collisions at the High-Luminosity LHC. In fact, ALPs have been searched at the LHC in several production and decay modes conserving flavor~\cite{dEnterria:2013zqi, CidVidal:2018blh, Goncalves:2021pdc,CMS:2018nsh, CMS:2019spf, ATLAS:2020ahi, CMS:2020ffa, ATLAS:2021hbr,ATLAS:2020pcy,dEnterria:2013zqi, CidVidal:2018blh, Goncalves:2021pdc,CMS:2018nsh, ATLAS:2020ahi, CMS:2020ffa, ATLAS:2021hbr,CMS:2020ffa,CMS:2018nsh, CMS:2019spf, ATLAS:2021hbr}, resulting in a variety of constraints~\cite{d'Enterria:2753504}. Previous phenomenological studies have anticipated the viability of these searches and others involving SM particles and new neutrinos in final states at the LHC, for example~\cite{Mimasu:2014nea,Brivio:2017ije,Bauer:2018uxu,Bauer:2017ris,Bauer:2017nlg,Jaeckel:2015jla,Alves:2016koo,Alves:2019xpc,Gavela:2019cmq,Merlo:2019anv,Baldenegro:2019whq,Dobrich:2015jyk,Harland-Lang:2019zur,Dobrich:2019dxc,Hahn:1404985,Alekhin:2015byh,Feng:2018pew,Aloni:2019ruo,Alves:2021puo,Knapen:2021elo,Alonso-Alvarez:2023wni,Cheung:2023nzg}. 

In this work, we explore the hypothesis that the dominant interactions of those new scalar particles violate flavor. If they do not couple diagonally in the theory's flavor space but maximally violate the flavor conservation, then there will be no coupling with same-flavor bosons as well; that is, there will be no relevant couplings to photons, gluons, and electroweak gauge boson pairs. \st We will denote the ALPs by \textit{flavons}, independently of the symmetry they are related. Of course, avoiding the hard constraints from flavor-changing neutral currents (FCNC) demands that flavor violation is predominantly a third-family effect, where flavons couple to top quarks. Our motivation thus is mainly phenomenological. As various experiments searching for flavons exclude models as they gather more data, the model building should follow the observations and adjust to them. Since no couplings of flavons to same-flavor fermions and gauge boson pairs have been observed, the hypothesis of dominant off diagonal fermion couplings becomes increasingly attractive. The effective field theory (EFT) formulation is particularly useful in this case once we do not need to tackle the intricacies of UV completions and leave the work open to further investigations. However, we anticipate that finding such a model is likely in frameworks like the Froggatt-Nielsen models, for example, where great freedom is given to build flavor couplings. In most models though, diagonal couplings will be present.

On the basis of an EFT, in which the flavons always change the flavor of SM fermions, we study their single and double production at the LHC with energy 13 and 14 TeV at the center of mass. These scalars couple to mass, so that their couplings are stronger with the top quark and the tau lepton. We just assume that the EFT originates from an ultraviolet completed theory containing an approximated global symmetry, whose spontaneous breaking of an energy scale $v\simeq\Lambda$ gives rise to the flavon. 

We show that flavons lighter than the top quark can be easily identified in proton-proton collisions at the LHC in scenarios where only off diagonal fermion couplings are present by performing a careful and systematic background estimation. In this mass range, the largest branching ratio is when the flavon decays into tau plus an electron or muon. For the pair production, $q\bar{q}^\prime\,\rightarrow aa$, clean final states comprising same-sign taus, \textit{i.e.} $\tau^{\pm}\tau^{\pm}+\ell^\mp\ell^\mp$, with $\ell=e,\mu$, arise forming $\tau^\pm\ell^\mp$ resonances. We also explore the associated production of flavon plus a top quark, $qg \,\rightarrow ta$, showing the discovery potential for both search channels when systematic uncertainties are considered. We find that double production presents larger sensitivity in most of the flavon mass range, but single production complements the search for masses close to the top mass. 

The impact of flavons on lepton flavor violations (LFV) has also been taken into account. We identify regions of the EFT parameter space that respect unitarity bounds, explain the ($g-2$) discrepancy of the muon  magnetic moment, and are safe from the current experimental constraints from the total width of the top quark. In particular, we find parameter instantiations of the maximally flavor violating model that escape the strong constraints from the LFV decays of the muon, $\mu\rightarrow\,e\,\gamma$, and the tau, $\tau\rightarrow\,e\,\gamma$, $\tau\rightarrow \mu\,\gamma$. Notwithstanding our working hypothesis, we analyze the impact of sizeable diagonal couplings in the fermion flavor space and with gauge boson pairs in the prospects of the LHC to observe the signals anticipating models where the presence of these couplings is present.

Such a flavon lighter than the top quark is also motivated by recent ATLAS data on top-quark distributions which cannot be explained by the SM event generators within the current experimental uncertainties~\cite{ATLAS:2023gsl,Banik:2023vxa}. A flavon exchanged in the $t$ channel might turn incoming up and charm quarks into top quarks distorting the SM distributions.  There are also hints from observed excesses in same-sign lepton signals~\cite{ATLAS:2022xnu} that could be explained by the double production of maximally violating flavons. These are hypotheses we plan to test in a future investigation.

Scalars with flavor-changing interactions have been searched for at the LHC. For example, the top-quark pair production with one of them decaying into a light quark plus a scalar that decays into a $b\bar{b}$ flavor-conserving pair~\cite{ATLAS:2023mcc} and the Higgs boson or additional Higgs bosons decays into $e^\mp\mu^\pm$~\cite{CMS:2023pte} and  $\tau^\pm\ell^\mp$~\cite{ATLAS:2019pmk}. The production signals at LHC of such particles were studied in top-quark decays from top-quark pair production~\cite{Carmona:2022jid,Bhattacharyya:2022umc}, gluon fusion and with a single top with decay modes into heavy quarks and leptons~\cite{Tsumura:2009yf}, in association with the $Z$ boson~\cite{Bhattacharyya:2022ciw},  and in decays with multilepton signatures~\cite{Hiller:2020fbu,Bissmann:2020lge}. Besides the collider phenomenology, there are also investigations in connection with limits on lepton flavor-violation decays of leptons, mesons,  discrepancies of the muon, and electron anomalous magnetic momentum~\cite{Cornella:2019uxs,Bauer:2019gfk,Carmona:2021seb,Muong-2:2006rrc,Muong-2:2015xgu,Muong-2:2023cdq,Hanneke_2008,Hanneke:2010au}. 

Our work is organized as follows: in Sec.~\ref{sec:model} we present the effective model and its parameters, on which our analysis is done; in Sec.~\ref{exp-const} the experimental constraints over the ALP are discussed in detail; in Sec.~\ref{backs} we discuss the signal of process and the relevant backgrounds; we present our analysis in Sec.~\ref{sec:analysis} and the results in Sec.~\ref{sec:results}; our conclusions are given in Sec.~\ref{conclusao}.

\section{Effective Model of Maximally Violating Flavons}
\label{sec:model}

The dimension-five effective Lagrangian describing flavon field, $a$, interactions is given by \cite{Cornella:2019uxs} 
\begin{equation}
    \mathcal{L} = - \frac{\partial_\mu a}{\Lambda} \sum_{f, i, j} \overline{f_i} \gamma^\mu (v_{ij} - a_{ij}\gamma_5) f_j \,,
    \label{eq:Lf}
\end{equation}
where $\Lambda$ is the new physics scale, and the axial and vectorial coupling constants between $a$ and the SM fermion flavors $f_i$ and $f_j$ are, respectively, $a_{ij}$ and $v_{ij}$. Through the use of the equations of motion, we can derive from Eq.~\eqref{eq:Lf} the following interaction terms of the flavon with fermions 
\begin{equation}
\label{eq:LFV-lagrangian}
    \mathcal{L} \supset -i \frac{a}{\Lambda} \sum_{f, i, j} \overline{f_i} \left[ (m_j - m_i) v_{ij} +  (m_j + m_i) a_{ij} \gamma_5 \right] f_j \,.
\end{equation}
In this form, we see that depending on the coefficients $v_{ij}$ and $a_{ij}$, the major contribution for couplings of the same order comes from the third family of fermions. The Lagrangian in Eq.~(\ref{eq:Lf}) might come from a theory with a global $U(1)_F$ spontaneously broken flavor symmetry under which each fermion can carry a different charge, which leads to the FV interactions. We also assume that this $U(1)_F$ symmetry is explicitly broken at the scalar potential giving rise to a mass for the flavon. Taking into account that $a$ is a pseudoscalar field, it can be seen that the FV couplings $v_{ij}$ imply violation of parity symmetry in Eqs.~(\ref{eq:Lf}) and (\ref{eq:LFV-lagrangian}). However, the CP symmetry is preserved if $v_{ij}$ and $a_{ij}$ are real. 

Diagonal fermion couplings induce interactions with photons, gluons, and SM weak bosons. All these interactions lead to several strong experimental constraints that must be avoided in the search for a viable model. Gauge boson couplings can be suppressed by assuming that the diagonal couplings to SM and new quarks and leptons are either absent from the UV complete theory or that the new particles are too heavy to compete with the nondiagonal SM fermions decays. In order to bypass those constraints, we turn off all the diagonal lepton and quark couplings to the flavon 
\begin{equation}
a_{ii} = 0 \,, \quad i=\ell, q.
\end{equation}

To avoid large FCNC at tree level, we only allow couplings of the top quarks to up and charm quarks. This way, the only nonzero parameters of our model are the following:
\begin{equation}
\left\{
    \begin{array}{cc}
m_a, \Lambda & \hbox{flavon mass and new physics scale} \\
v_{tc}, a_{tc}, v_{tu}, a_{tu} & \hbox{non-diagonal top-quark couplings} \\
v_{e\tau}, a_{e\tau}, v_{\mu\tau}, a_{\mu\tau}, v_{e\mu}, a_{e\mu} & \hbox{non-diagonal lepton couplings}
    \end{array}
    \right.
\end{equation}

Couplings to bottom-strange and bottom-down quarks can, however, be generated at the one-loop level involving the top-charm or top-up quark couplings. Because of the strong experimental suppression of the tree-level coupling between the bottom quark and the flavon,  the loop-induced coupling is effectively a next-to-next-to-leading order electroweak contribution which turns out to be, at least, 2 orders of magnitude smaller than the tree-level $a\to \tau+\ell$ and can be safely neglected for our purposes; see Appendix~\ref{app:loop}. Notice, by the way, that if those contributions were sizeable, strong indirect FCNC limits could be placed on the top-quark couplings to flavons.

In this work, we will restrict our analysis to flavons lighter than top quarks, with masses between 10 and 170 GeV\footnote{Maximally violating flavons heavier than the top quark will be discussed in a forthcoming work.}. As a consequence,  the flavon branching ratio is almost 100\% in taus plus an electron or muon and only a tiny fraction into electron muon. Note that producing these scalars at hadron colliders is viable only through top-quark couplings. Because diagonal couplings to fermions are absent, effective couplings to all SM bosons are suppressed. Therefore, gluon fusion, weak boson fusion, and associated production with weak bosons are out of hand at hadron colliders.

The Standard Model is a consistent theory and does not violate unitarity at high energies. However, when we add higher dimensional effective operators, the theory might predict scattering cross sections that grow with the center-of-mass energy of the process violating the optical theorem at some point. For our Lagrangian, the strongest bounds from the $f_i \bar{f}_j \rightarrow f_i \bar{f}_j$ scattering are given by
\begin{equation}\label{eq:unitarity-bounds}
  \left|v_{ij} \right| < \sqrt{8\pi}\frac{\Lambda}{m_j\,-\,m_i} \;\;\mbox{and}\;\; \left|a_{ij} \right| < \sqrt{8\pi}\frac{\Lambda}{m_j\,+\,m_i}\, .
\end{equation}
The axial and vector couplings can be as large as $\left|v_{3i}\right|/\Lambda$, $\left|a_{3i}\right|/\Lambda \lesssim 1.4\times 10^{-2}(1.5)$ GeV$^{-1}$ in the case of top(tau) quark(lepton) couplings. These bounds differ somewhat from those of Ref.~\cite{Cornella:2019uxs} because, contrary to that work, we are assuming only nondiagonal couplings; see Appendix~\ref{appendix:unitarity}.

\section{Experimental Constraints}
\label{exp-const}
 
 Besides the partial wave unitarity bounds, many experimental constraints apply. For the mass range we consider, we need to check the flavon contribution to the decay of top quarks, $Z$, and $W$ bosons and LFV-induced transitions. We list below the constraints considered in this work.

\begin{enumerate}
\item The measured model-independent top-quark decay width of $1.9 \pm 0.5$ GeV~\cite{ParticleDataGroup:2022pth} imposes an upper limit on the top-quark-flavon coupling. The top-quark partial width into $Wb$ is measured to be $1.42^{+0.19}_{-0.15}$ GeV~\cite{ParticleDataGroup:2022pth}. Assuming the difference from this value to the experimental constraint is due to the new channels involving the new scalar, we conservatively impose the constraint
\begin{equation}
\Gamma(t\to a+u)+\Gamma(t\to a+c) < |1.9-1.4|=0.5 \, \hbox{GeV}\, . 
\label{eq:br}
\end{equation}
The analytic expression for the width $t \to aq$, with $q = u,c$ is given in Appendix \ref{app:decay}. This constraint implies $\mathrm{BR}(t\to aq)<0.25$, $q=u,c$.

\item The lepton-flavon coupling contributes to the $Z$ total decay width if its mass is such that $Z\to \tau+\ell+a$ is allowed. We checked that the new partial width is of order $10^{-11}$ GeV in the flavon mass range up to $m_Z-m_\tau$, thus much smaller than the $Z$ boson width uncertainty of $10^{-3}$ GeV~\cite{ALEPH:2005ab}. We also checked that the impact of flavons on the $W$ boson width~\cite{ALEPH:2010aa} is negligible.

\item It is also necessary to check if the flavon decays fast enough for its by-products to be found inside the detector. For the mass range and couplings of interest in this work, we found that the flavon decays promptly after its production well inside the inner tracking region of the detector.

\item The most stringent constraint comes from the MEG experiment of muon decay to electron and photon \cite{MEG:2016leq}. Following the null results,  the limits on the branching ratio of the muon in this channel is of order $10^{-13}$. The planned MEG II is expected to improve that limit by an order of magnitude. The BABAR Collaboration, by its turn, searched for this LFV process in the tau-channel; the limits in $\tau \rightarrow e \gamma$ and $\tau \rightarrow \mu\gamma$ are of order $10^{-8}$ \cite{BaBar:2009hkt}. 

\end{enumerate}

\begin{figure}[t]
    \centering
       \begin{tikzpicture}[baseline=(current bounding box.center)]
    \begin{feynman}
        \vertex (a2);
        \vertex [right=of a2] (a3) {\(a\)};
        \vertex [left=of a2] (a1) {\(q\)};
        \vertex [below=of a2] (b2);
        \vertex [right=of b2] (b3) {\(t\)};
        \vertex [left=of b2] (b1) {\(g\)};
        \diagram* {
        (a1) -- [fermion] (a2) -- [fermion, edge label=\(t\)] (b2) -- [gluon] (b1),
        (a2) -- [scalar] (a3),
        (b2) -- [fermion] (b3)
        };
    \end{feynman}
\end{tikzpicture} \begin{tikzpicture}[baseline=(current bounding box.center)]
    \begin{feynman}
        \vertex (b2);
        \vertex [above left=of b2] (a1) {\(q\)};
        \vertex [below left=of b2] (c1) {\(g\)};
        \vertex [right=of b2] (b3);
        \vertex [above right=of b3] (a4) {\(a\)};
        \vertex [below right=of b3] (c4) {\(t\)};
        \diagram* {
            (a1) -- [fermion] (b2) -- [gluon] (c1),
            (b2) -- [fermion, edge label=\(q\)] (b3),
            (a4) -- [scalar] (b3) -- [fermion] (c4),
        };
    \end{feynman}
\end{tikzpicture}  \begin{tikzpicture}[baseline=(current bounding box.center)]
    \begin{feynman}
        \vertex (a2);
        \vertex [right=of a2] (a3) {\(a\)};
        \vertex [left=of a2] (a1) {\(q\)};
        \vertex [below=of a2] (b2);
        \vertex [right=of b2] (b3) {\(a\)};
        \vertex [left=of b2] (b1) {\(\bar{q}^{\prime}\)};
        \diagram* {
        (a1) -- [fermion] (a2) -- [fermion, edge label=\(t\)] (b2) -- [fermion] (b1),
        (a2) -- [scalar] (a3),
        (b2) -- [scalar] (b3)
        };
    \end{feynman}
\end{tikzpicture}
    \caption{Feynman diagrams for single (left and central panels) and double (right panel) production of maximally flavor-violating scalars in hadron collisions. Single production is suppressed by $(m_t/\Lambda)^2$ while double production is suppressed by $(m_t/\Lambda)^4$.}
    \label{fig:prod}
\end{figure}
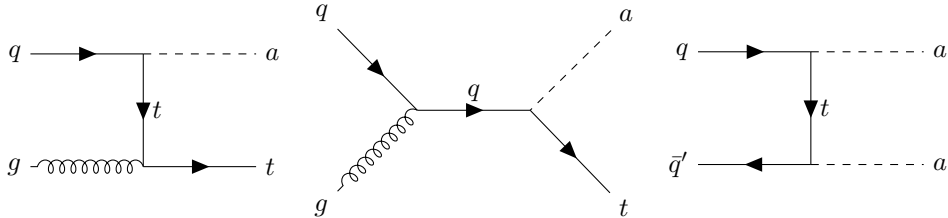

\section{Signal and Background Simulations}\label{backs}

We simulated signals and backgrounds in leading order at the 13 and 14 TeV LHC with \texttt{MadGraph5}~\cite{Alwall:2014hca, Frederix:2018nkq} using the \texttt{UFO}~\cite{Degrande:2011ua} files generated by \texttt{FeynRules}~\cite{Alloul:2013bka} where we implemented our effective model.  The parton shower and detector simulations were performed in the default settings of \texttt{PYTHIA 8.3} \cite{Bierlich:2022pfr} and \texttt{Delphes3}~\cite{deFavereau:2013fsa}, respectively. The signal we consider comprises simple and double flavon productions in Figure~\ref{fig:prod}. The branching ratio of the flavon to $\tau+\ell$ is nearly 100 \% if $m_a<m_t$, as discussed earlier. The signal events were simulated setting $a_{tq}=v_{tq}=1$, $q=u,c$ and $\Lambda=1$ TeV. With these choices, the width $\Gamma(t \to aq)$ described in Eq. \eqref{eq:top-decay} can be simplified and the signals are proportional to powers of $c_{tq} m_t/\Lambda$, where $c_{tq} = \sqrt{a_{tq}^2 + v_{tq}^2} = \sqrt{2}$. We rescale our signal cross sections generated with $\Lambda=1$ TeV, for a given fixed flavon mass, according to the value of $\Lambda$. We employed the \texttt{NN23LO1} PDF set~\cite{Ball_2017} in all simulations. In order to estimate the impact of higher order QCD corrections, we simulate signals and backgrounds with up to two extra jets in the Michelangelo L. Mangano merging approach~\cite{Mangano:2006rw}.

The main production modes of maximally violating flavons are double and single production.
The double production
\begin{equation}
    q\bar{q}^\prime \to aa\; ,
\end{equation}
where $q(q^\prime)=u,c$ with a top quark exchanged in the $t$ channel, is suppressed by $(m_t/\Lambda)^4$, but it is initiated by valence up quarks; see Figure \ref{fig:prod}, right panel. 

The single production where the flavon is produced in association with a top quark 
\begin{equation}
    qg\to at\; ,
\end{equation}
occurs via a $t$ channel diagram as well, as shown in the left and central panels of Figure \ref{fig:prod}. In this case, however, the final state is heavier compared to $a+a$ due to the presence of a top quark, but its cross section is suppressed by $(m_t/\Lambda)^2$. We also take the $\bar{t}+a$ contribution in our analysis.

\begin{figure}
    \centering
    \includegraphics[width=0.45\linewidth]{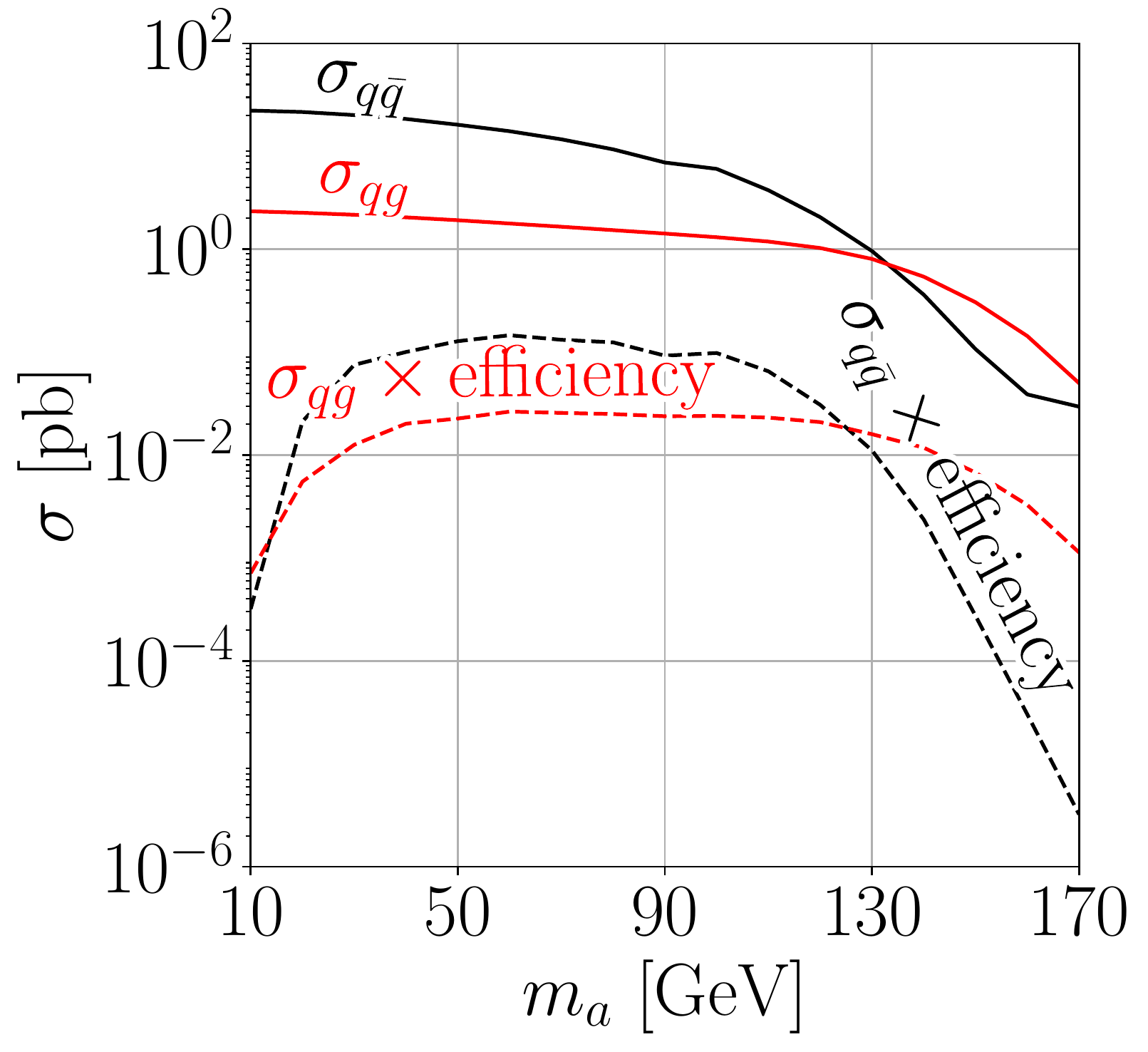}
    \includegraphics[width=0.45\linewidth]{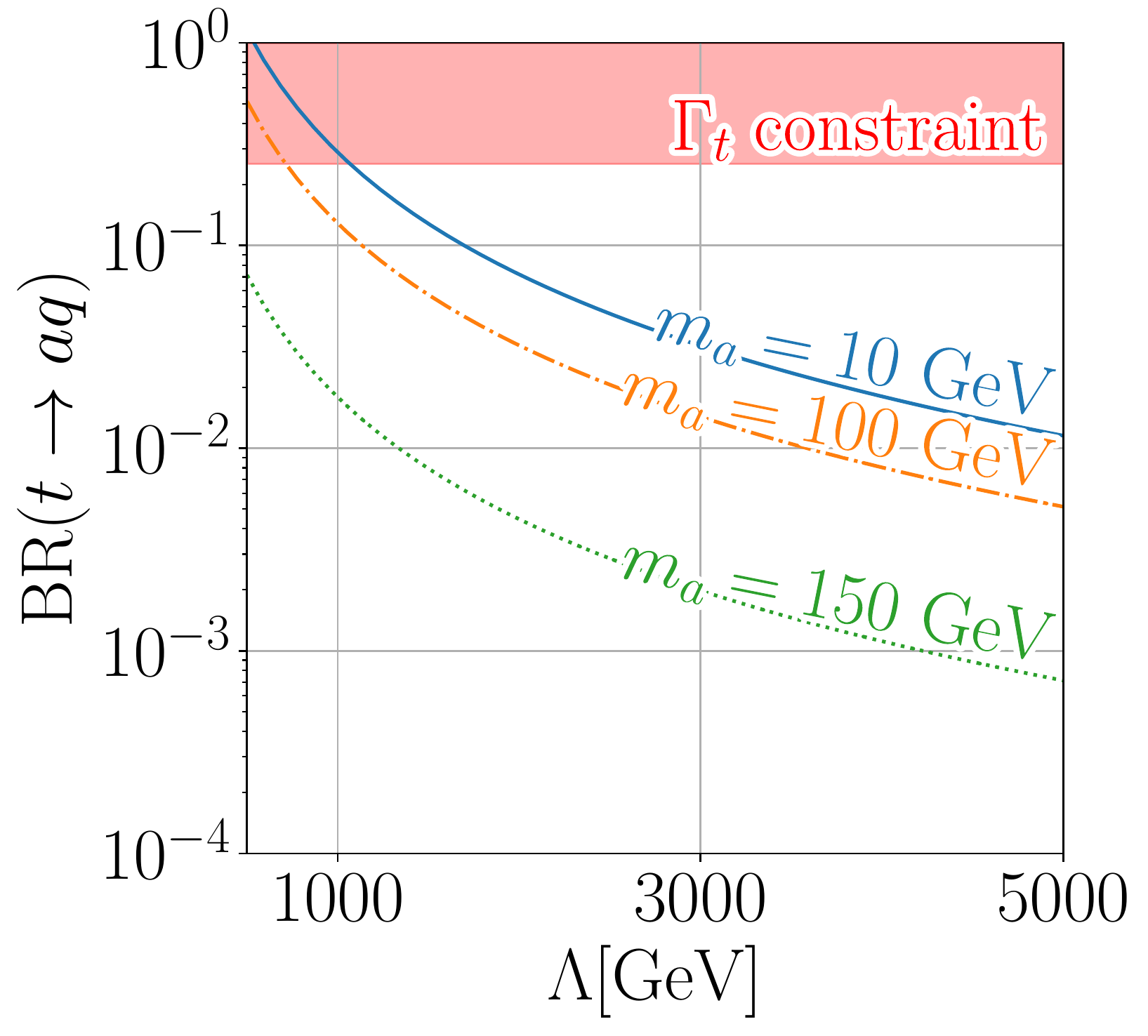}
    \caption{The production cross section of flavon pairs ($\sigma_{q\bar{q}}$) and associated flavon+top quark ($\sigma_{qg}$) at the 14 TeV LHC. The dashed lines depict the cross-section times selection efficiency as discussed in the text. The branching ratio BR$(t \to aq)$ for various values of flavon masses as a function of the cutoff scale, $\Lambda$. The shaded area represents the constraint from Eq.~\eqref{eq:br}.}
    \label{fig:xsec-times-efficiency}
\end{figure}

 We show, in Figure~\ref{fig:xsec-times-efficiency}, the left panel, the production cross section, in pb, for double (solid black line) and single (solid red line) production at the 14 TeV LHC. For masses up to $\sim 120$ GeV, the flavon pair production ($q\bar{q}^\prime$ channel) exceeds the single production ($qg$ channel), especially for lighter flavons. The combined effect of a lighter final state and two initial-state up quarks gives the double production an advantage even being suppressed by $(m_t/\Lambda)^4$. That advantage is lost as the flavons get heavy. The production cross section for both processes is larger than 1 pb for masses up to 120 GeV and drops to around 10 fb for $m_a\sim m_t$. These rates drop even further (dashed lines in Figure \ref{fig:xsec-times-efficiency}, left panel) as we impose selections and kinematic cuts on the final state particles as we discuss ahead.

Owing to the flavor violation, the double flavon production gives rise to same-sign $\tau$ leptons plus same-sign electrons/muons half the time. The taus decay to ``one-prong" signatures with one charged particle around 85\% of the time, while ``three-prong" decays with three charged particles occur at approximately 15\%. Total leptonic decays with one charged electron or muon have a branching ratio of 35.1\%, but leptonic decays make $\tau$ lepton identification more difficult. On the other hand, hadronic decays give rise to low multiplicity jets that can be more easily tagged as $\tau$ jets. Our strategy relies on the identification of charged $\tau$ leptons, so we propose a search for signals with hadronic taus 
\begin{equation}
q\bar{q}^\prime\to aa\to \tau_h^\pm\tau_h^\pm+\ell^\mp \ell^\mp,\; \ell=e,\mu\; .
\end{equation}
This is a very clean signature if we are able to reconstruct the $\tau$ leptons charges. Also, notice that the decay of taus yields neutrinos, and then the signals involve missing energy as well.

In the case of associated flavon+top quark production,  if the top quark decays to a flavon plus an up or charm quark, the two flavons generate the same-sign tau pair again, but the $t\to aq$ branching is suppressed by $\Lambda$ as shown in Figure~\ref{fig:xsec-times-efficiency}, right panel. Instead, we look at the dominant channel where the top quark decays to $Wb$ and still select events with same-charge taus with one tau originating from the $W$-boson decay
\begin{equation}
    qg \to a t \to a+W b\to \tau_h^{\pm} \tau_h^{\pm} + \ell^{\mp} + \nu_\tau + b \;.
\end{equation}

We show, in Tables~\ref{tab:double} and \ref{tab:simple}, all the reducible and irreducible background sources considered in our work. Next, we discuss the phenomenological analysis of the work.

\section{Analysis}\label{sec:analysis}

We adopt two selection strategies, one for each production mode. The selection for double production demands two same-charge taus, and two same-charge electrons and/or muons but opposite to the tau charge.  For the single production, events must also contain two same-charge taus, one $\ell$ with the opposite charge to the taus, and one $b$-tagged jet.  In both cases, we also impose the following basic cuts: %
\begin{eqnarray}
    && p_T^{\tau_h}  > 10 \, \si{GeV} \,, \left| \eta_{\tau_h} \right| < 2.5, \nonumber \\
    && p_T^{\ell}  > 10 \, \si{GeV} \,, \left| \eta_{\ell} \right| < 2.5, \nonumber \\
    && p_T^{j} >  30 \, \si{GeV} \,, \left| \eta_j \right| <  2.5, 
    \label{eq:basic_cuts}
\end{eqnarray}
where $\ell = e, \mu$ and $j$ denotes a $b$-tagged jet.

 The cross section times selection efficiency is shown in Figure~\ref{fig:xsec-times-efficiency}, the dashed lines at the left panel, for the two flavon production modes. Overall, the efficiency of identifying two tau jets plus leptons and/or $b$ jets is low. With a fixed efficiency for a hadronic tau of 0.6, the \texttt{Delphes3}~\cite{deFavereau:2013fsa} default value, and taking the branching ratio of tau into hadrons as 0.65, we have an initial efficiency of $(0.65\times 0.6)^2=0.15$ for two hadronic taus. Taking into account minimal requirements for triggers, generation cuts, and identification of leptons, that efficiency drops further. However, we profit from almost vanishing background rates in these channels.

\begin{table}[t]
\scalefont{1}

\begin{tabular}{c c}
\hline
Background & $\sigma_{\text{basic}}$ (fb) \\ 
\hline\hline

$t\bar{t}\to W^\pm W^\mp b\bar{b}\to \tau^{\pm}_h \ell^{\mp} + b\bar{b}+\nu\nu$ & $<0.5366$ \\ 
\hline

$tW^\pm \to W^\mp W^\pm+b\to \tau^{\pm}_h \ell^{\mp} + b+ \nu\nu$ & $<0.0604$ \\ 
\hline 

$W^\pm W^\mp\to \tau_h^\pm + \ell^\mp + \nu\nu$ & $<0.06694$   \\  

$ZZ\to \tau_h^\pm\tau_{\ell}^\mp + \ell^\pm\ell^\mp$ & $0.0037$    \\ 

$ZZ \to 4\tau \to \tau_h^\pm \tau_\ell^\pm \tau_h^\mp \tau_\ell^\mp + 4 \nu$  & $5.4 \times 10^{-21}$\\
$ZW^\pm \to \tau^+ \tau^- + \ell^\pm + \nu$ & $0.012$ \\  

$ZW^\pm\to \ell^+\ell^- + \tau_h^\pm + \nu$  & $<0.003978$ \\ 
\hline

$W^+W^-Z(\gamma^*)\to \tau^+\tau^{-}\tau^{\pm}_{h}\tau^\mp_\ell + \nu\nu$  & $8.5 \times 10^{-4}$ \\  
\begin{tabular}[c]{@{}c@{}}$W^\pm W^\mp W^\pm \to \tau^{\pm}_h\tau^{\pm}_h + \ell^\mp + \nu\nu\nu $\end{tabular}                                                         & $<6.7 \times 10^{-7}$           \\ 
\begin{tabular}[c]{@{}c@{}}$W^+W^-Z\to \tau^{\pm}_h\tau^{\mp}_\ell\tau^{\pm}_h + \ell^\mp + \nu\nu $\end{tabular}   & $7.4 \times 10^{-11}$    \\  
$W^+W^-\gamma^*\to \tau^{\pm}_h\tau^{\mp}_\ell\tau^{\pm}_h+ \ell^\mp + \nu\nu$                                    & $2.9 \times 10^{-7}$    \\ \hline 
Total background & $< 0.684$ \\
\hline
\end{tabular}
\centering
\caption{Reducible and irreducible background cross sections, in fb, for the double production after the selection efficiencies at 14 TeV LHC. When a leptonic tau appears, it decays to $\tau^{\mp}_\ell\to \ell^\mp+\nu\nu$. Tau leptons without a subscript can be either of leptonic or hadronic type. In the last line of the table, we calculate the superior limit of the total background rate. This limit is taken to obtain the statistical significances after the cut on the tau-lepton mass as discussed in the text.}
\label{tab:double}
\end{table}
 Selecting same-sign taus and leptons is advantageous from the point of view of background contamination. In the SM, the dominant irreducible source for double production is $ZZ\to \tau_\ell^\pm\tau_h^\mp+\tau_\ell^\pm\tau_h^\mp$~, where two taus ($\tau_\ell$) decay leptonically and the other two ($\tau_h)$, hadronically.   The $\tau_h^\pm\tau_h^\pm+\ell^\mp\ell^\mp$ final state has been investigated in the scope of the 2HDM~\cite{Iguro:2019sly,Blanke:2022kpi} and the SM $ZZ\to 4\tau$ background was estimated to be negligible. However, many reducible background sources might contribute to backgrounds. We list in, Table~\ref{tab:double}, the background sources we consider in our analysis for the double flavon production and decay. The common feature of those background processes is the presence of additional leptons/$\tau$ jets from proton fragmentation and QCD radiation, which, combined with produced leptons/hadronic taus, might mimic the flavon pair signature. As anticipated in Refs.~\cite{Iguro:2019sly,Blanke:2022kpi}, the $ZZ$ background is negligible after selecting two same-sign taus and two same-sign leptons since the same-sign leptons originate from suppressed leptonic tau decays. On the other hand, we estimate that reducible backgrounds amount to approximately 0.7 fb, at most, with reducible $t\bar{t}$, $tW$, and $WW$ as the major contributions. As discussed in the next section, these backgrounds can further be suppressed by identifying the $\tau^\pm\ell^\mp$ peaks in their invariant mass distributions.

The background sources for single production are shown in Table~\ref{tab:simple}. In this case, we must also deal with $t\bar{t}$ and $tW$ production. However, as the final state of interest has just one charged electron/muon, contrary to the double production, the backgrounds pass the selection requirements more easily. Our estimate for the upper limit of the total cross section is much larger now, reaching around 65 fb. In Tables~\ref{tab:double} and \ref{tab:simple}, the inequalities indicate an upper limit estimate of the cross sections where no Monte Carlo events were found after imposing selections and basic cuts of Eq.~\eqref{eq:basic_cuts}.

\begin{table}[h!]
\scalefont{1}
\begin{tabular}{c c}
\hline
 Background & $\sigma_{\text{basic}}$ (fb) \\ \hline\hline 
 $t\bar{t}\to W^\pm W^\mp b\bar{b}\to \tau_{h}^\pm \ell^\mp + b\bar{b}+\nu\nu$ & $47.82$\\ 
 
 $t\bar{t}\to W^\pm W^\mp b\bar{b}\to \tau_{h}^\pm \tau_{h}^\mp + b\bar{b}+\nu\nu$ & $4.88$ \\ 
 \hline

 $tW^\pm \to W^\mp W^\pm+b\to \tau_{h}^\pm \ell^\mp + b+ \nu\nu$ & $2.746$       \\ 
 $tW^\pm \to W^\mp W^\pm+b\to \tau_h^\pm \tau_{\ell}^\mp + b+ \nu\nu$ & $0.347$ \\ \hline
 
 $ZW^\pm(j)  \to \tau^+\tau^- + \tau_{h}^\pm + \nu$ & $0.02113$ \\ 

 $W^\pm W^\mp(j) \to \tau_h^\pm + \ell^\mp + \nu\nu$ & $0.1029$ \\ 
 
 $W^\pm W^\mp(j) \to \tau_h^\pm + \tau_{\ell}^\mp + \nu\nu$ & $<0.0129$ \\ 
 \hline

 $W^\pm W^\mp W^\pm(j)  \to \tau_h^\pm\tau_h^\pm + \ell^\mp + \nu\nu\nu$ & $6.92\times 10^{-4}$  \\ 
 
 $Z Z W^\pm \to b \bar{b} + \tau^+ \tau^- + \ell^{\pm} \nu$ & $1.14 \times 10^{-4}$ \\ 
 \hline

 $Z j \to \tau_h^{\pm}\tau_{\ell}^{\mp}j$ & $<8.64$ \\ 
 \hline
 Total background & $<64.58$ \\
 \hline
\end{tabular}
\caption{Background processes for the associate top+flavon production at the 14 TeV LHC. In the last line of the table, we calculate the superior limit of the total background rate. This limit is taken to obtain the statistical significances after the cut on the tau-lepton mass, as discussed in the text. Tau leptons without a subscript can be either of leptonic or hadronic type.}
\label{tab:simple}
\end{table}
We use the Asimov estimate for the statistical significance~\cite{Cowan_2011}
\begin{equation}\label{eq:asimov}
    Z(S, B, \sigma_B^2) = \sqrt{2 (S+B)\ln \frac{(S+B)(S+\sigma_B^2)}{B^2 + (S+B)\sigma_B^2} - 2 \frac{B^2}{\sigma_B^2} \ln \left(1 +\frac{S \sigma_B^2}{B^2 + B \sigma_B^2}\right)} \,,
\end{equation}
where $S$ and $B$, are the number of surviving signal and background events after the cuts are performed, and $\sigma_B^2 = \varepsilon_B B$ is uncertainty in the measure of $B$ as a function of the systematic uncertainty of the total background rate, $\varepsilon_B$. When no background events survive, we estimate the significance as $Z=\sqrt{S}$.

\begin{figure}[t]
    \centering
    \includegraphics[width=0.45\linewidth]{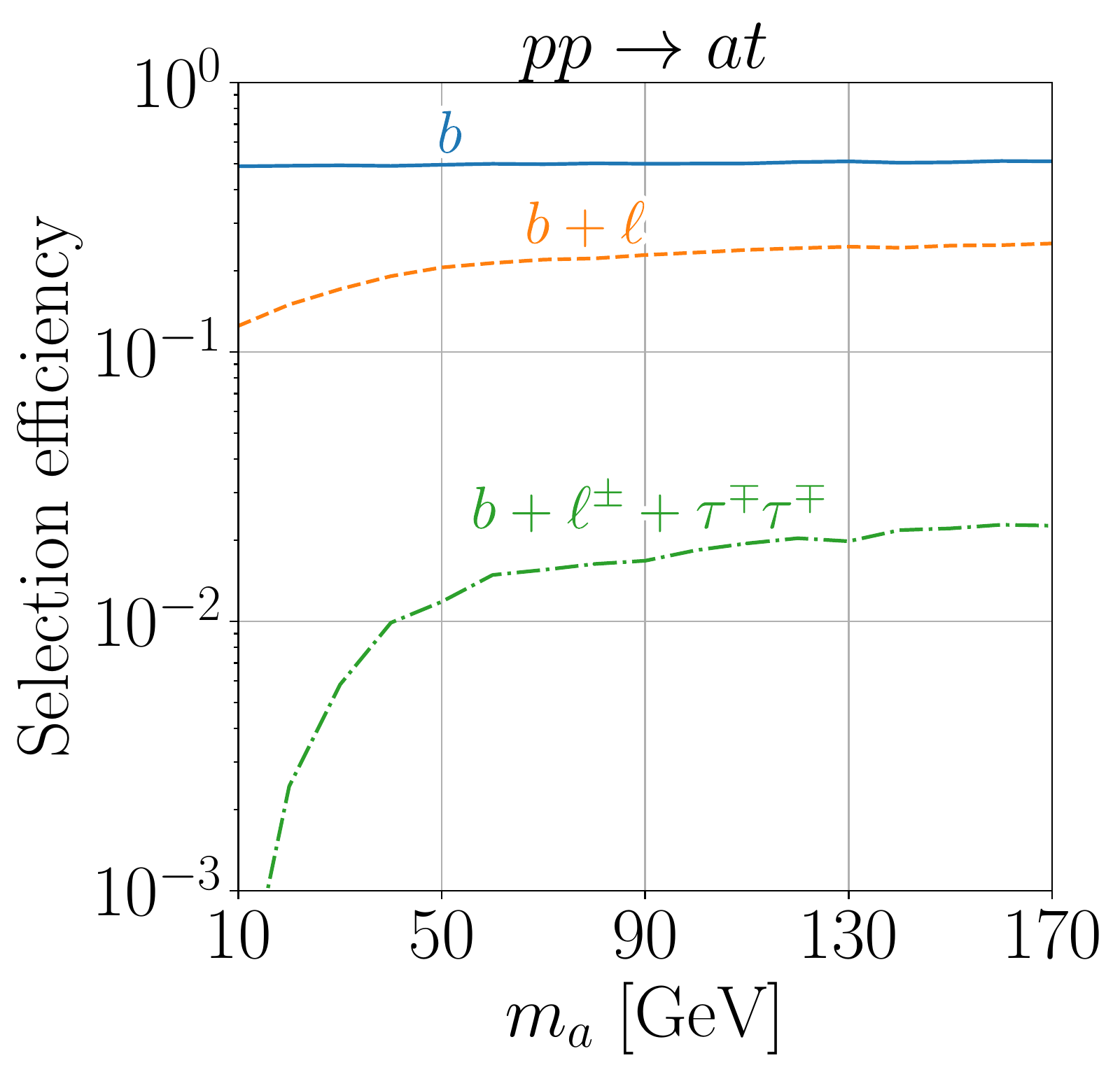}
    \includegraphics[width=0.45\linewidth]{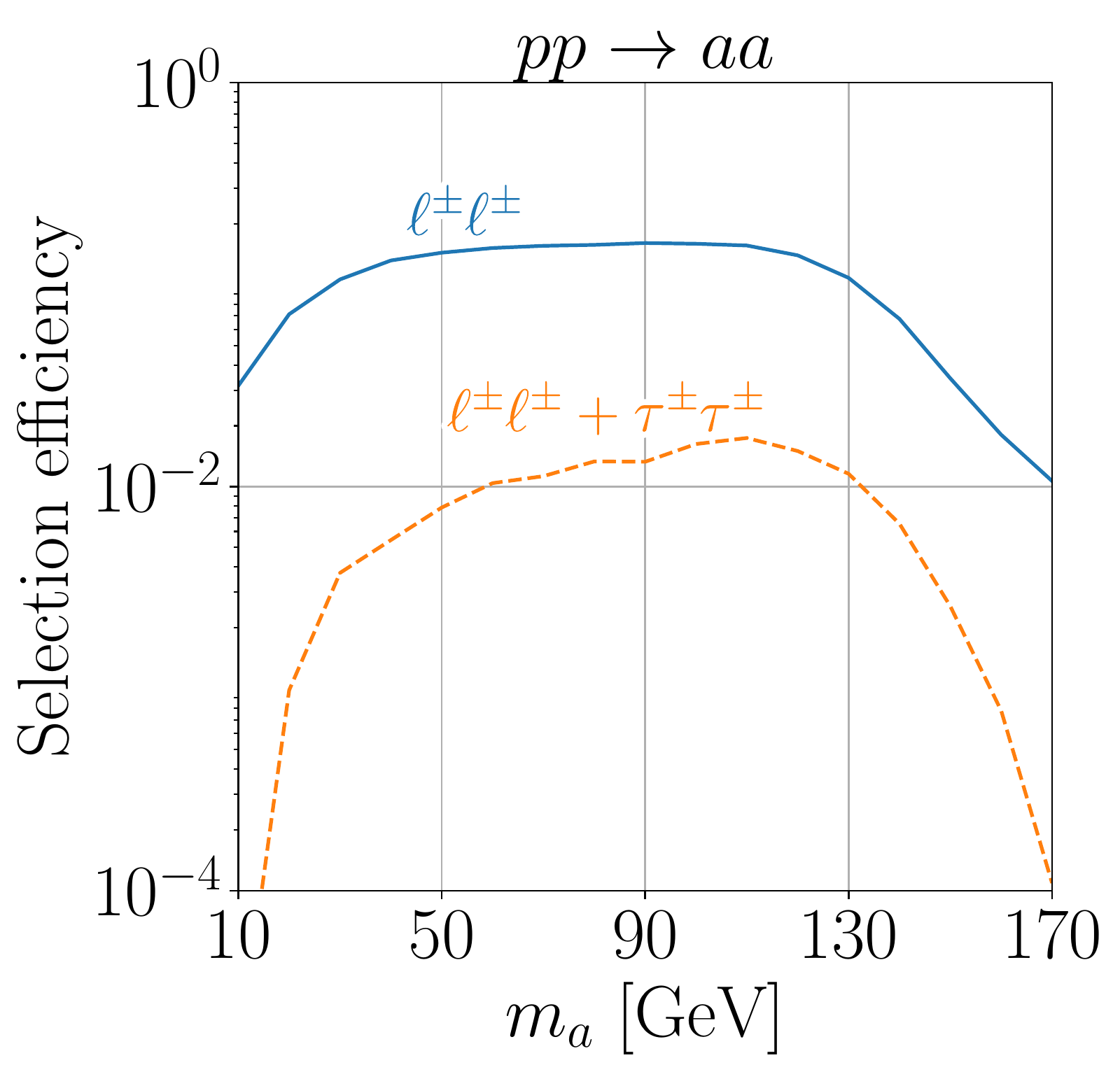}
    \caption{Left: the percentage of surviving events after successive final state selections in the associate $a+t$ production. Right: the selection efficiencies for double production. In both cases, requiring two taus with the same charge has the biggest impact on the overall identification efficiency.}
    \label{fig:id}
\end{figure}
In Figure \ref{fig:id}, we depict the signal selection efficiency when we impose the sequential identification  of particle species. In the right panel, the identification of two same-sign tau jets in double production occurs in 10\% of the events with two identified same-sign electrons/muons in the 60--140 GeV mass range, approximately, but drops to only 1\% of $\ell^\pm\ell^\pm$ for $m_a<60$ GeV and flavon masses close to the top-quark mass. The maximum overall selection efficiency reaches $\sim 0.01$ for $m_a\sim 120$ GeV. In the case of associate flavon-top quark production, identifying events with at least one $b$ jet is efficient, and identifying at least one $b$ jet and one charged electron/muon reaches 20\% for heavier masses. However, as in the double production case, events with two additional tau jets of same sign are much rarer. The overall identification efficiency also reaches 0.01 at most, but, different from the double production, the efficiency increases with the flavon mass toward the top mass. Despite the low signal identification efficiency, the impact on the backgrounds is much more severe, and choosing such a channel pays off.
In order to optimize the statistical significance even further, we impose a cut on the highest $p_T$ pair $\tau^\pm_h\ell^\mp$ mass, $M_{\tau\ell}$, determined by $|M_{\tau\ell}-m|\leq \delta m$, and performing a random search on $m_a$ and $\delta_m$ such that
\begin{equation}
    N_\sigma = \underset{m_a, \delta m}{\text{argmax}} \, Z(S(m_a, \delta m), B(m_a, \delta m), \varepsilon_B)\; . 
\end{equation}

\begin{figure}[t]
    \centering
    \includegraphics[width=0.45\linewidth]{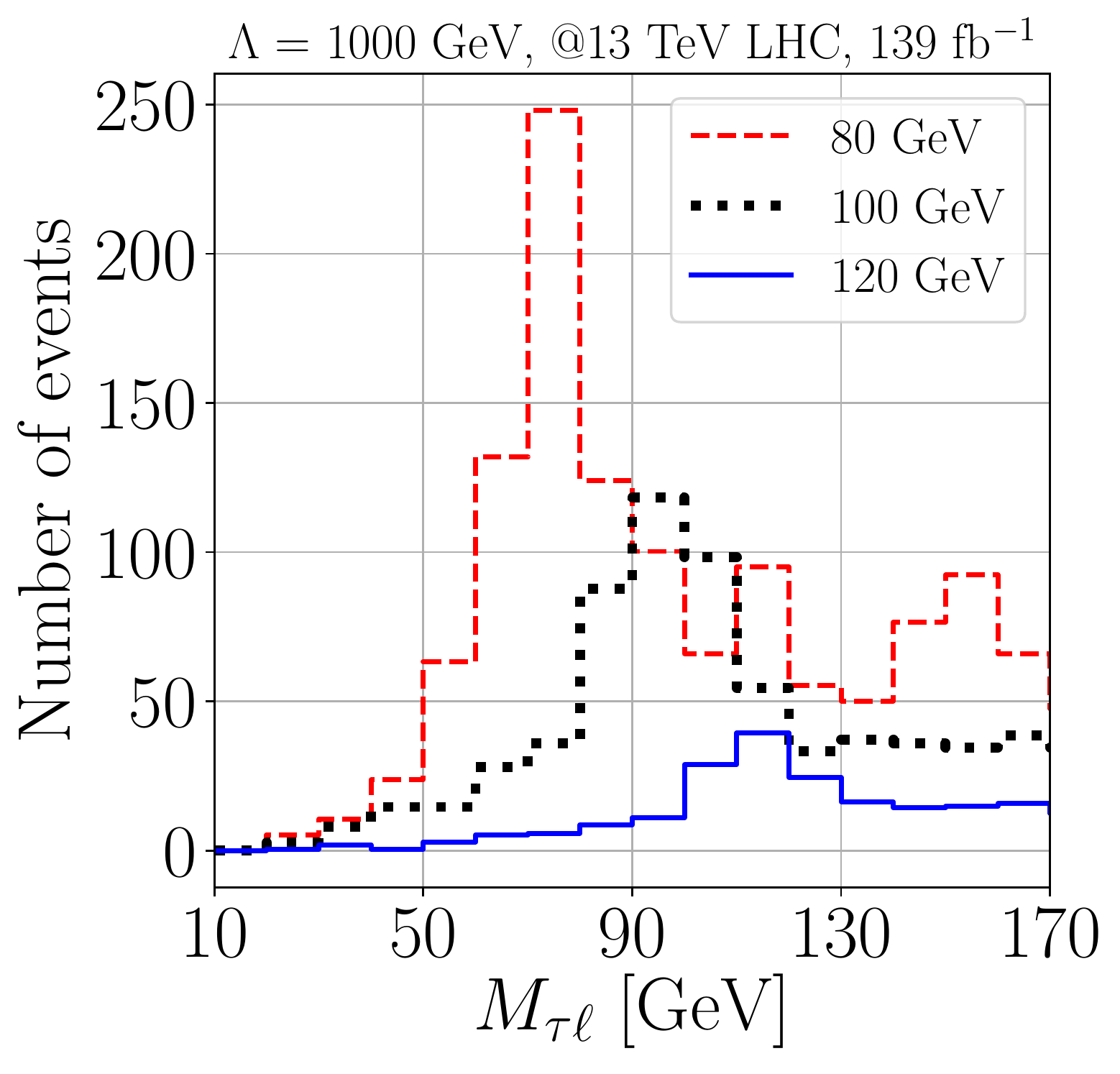}
    \caption{The $\tau^\pm$-$\ell^\mp=e,\mu$ invariant mass for the double production $q \bar{q}^\prime \to aa$ for flavon masses of 80, 100, and 120 GeV. The histogram is normalized by the number of events expected at the 13 TeV LHC after 139 fb$^{-1}$. We estimate a negligible number of background events in this channel.}
    \label{fig:mtaulep}
\end{figure}

Here $S(m_a,\delta m)$ and $B(m_a,\delta m)$ represent the number of signal and background events after the cut on the mass of the $\tau$-lepton pair for fixed $\varepsilon_B$. This cut turns the backgrounds to double production negligible and reduces the backgrounds to single flavon production by a factor of $\sim 0.25$ for lighter masses raising to no reduction at all for larger masses. We show, in Figure~\ref{fig:mtaulep},  the $\tau^\pm\ell^\mp$ mass for the double production process, after 139 fb$^{-1}$ of integrated luminosity at the 13 TeV LHC, in the cases where the flavon has masses of 80, 100, and 120 GeV and $c_{tq}/\Lambda=1$ TeV$^{-1}$ prior to the invariant mass cut. In the absence of backgrounds, these peaks should be easily identifiable; however, our reconstruction of the invariant mass is simplified compared to experimental analyses. For example, the search for LFV SM Higgs boson decays to $\tau^\pm\ell^\mp$~\cite{ATLAS:2019pmk} comprises multivariate tools to identify the various objects in final states with QCD and $\tau$ jets. Nonetheless, a reconstruction of the scalar resonance should be feasible within some degree of resolution.

\section{Results}
\label{sec:results}

 We now present a sensitivity analysis for the potential of the 13 and 14 TeV LHC to exclude or discover the proposed signals of the model in the flavon mass range from 10 to 170 GeV.  

 \subsection{Sensitivity of the LHC to top-quark couplings and flavon masses}
\begin{figure}[t]
        \centering
        \includegraphics[width=0.45\linewidth]{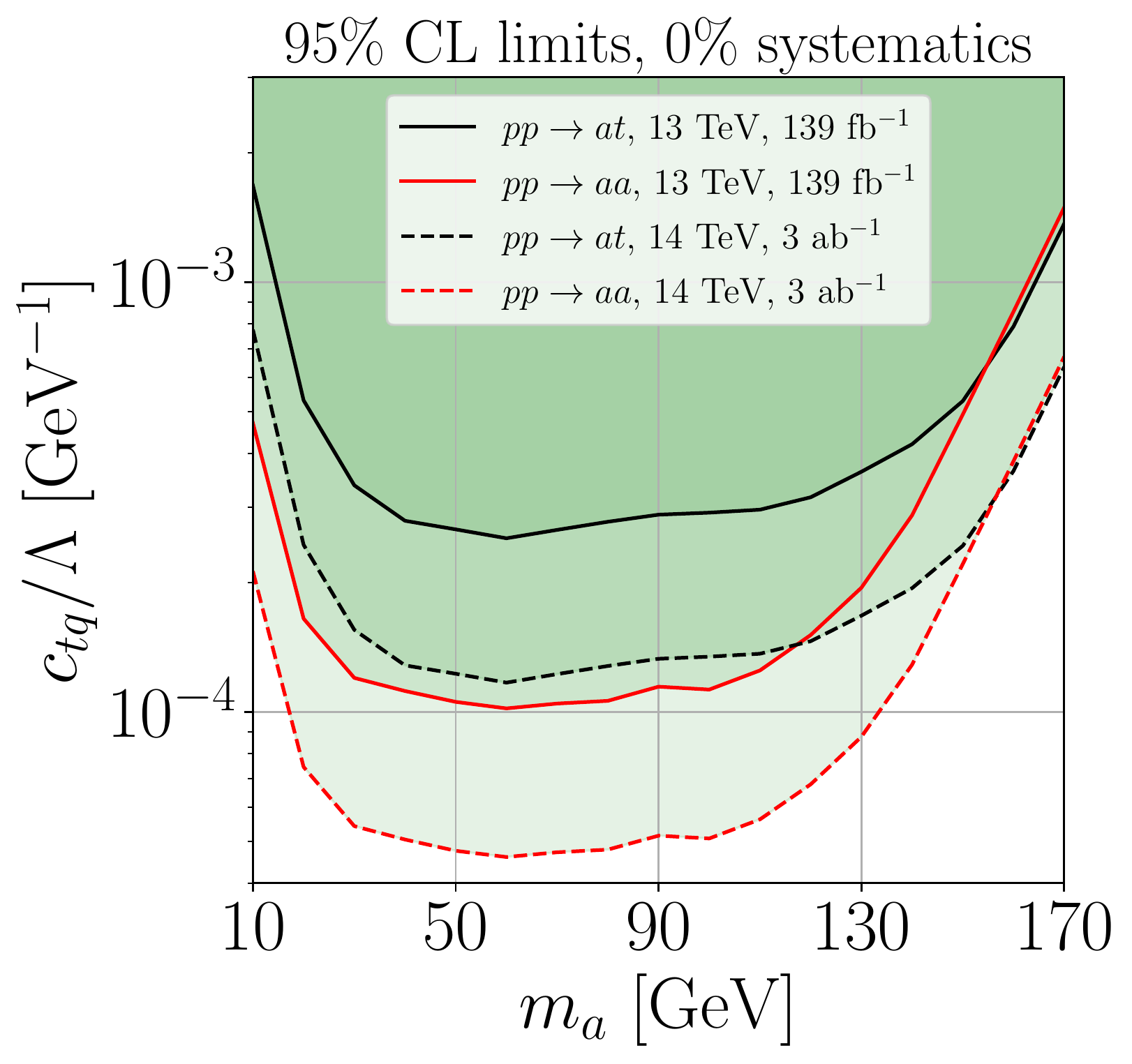}
        \includegraphics[width=0.45\linewidth]{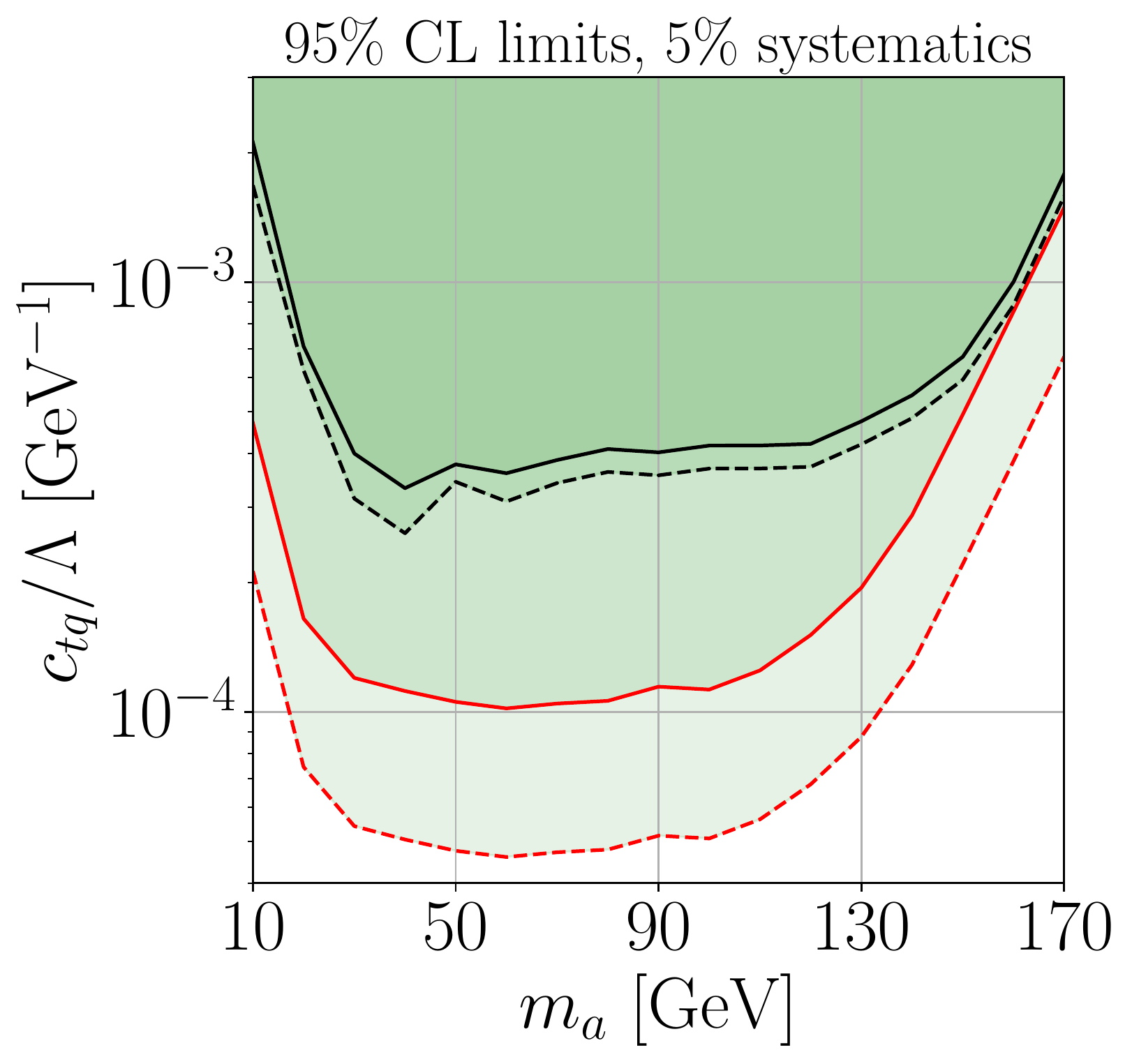}
        \includegraphics[width=0.45\linewidth]{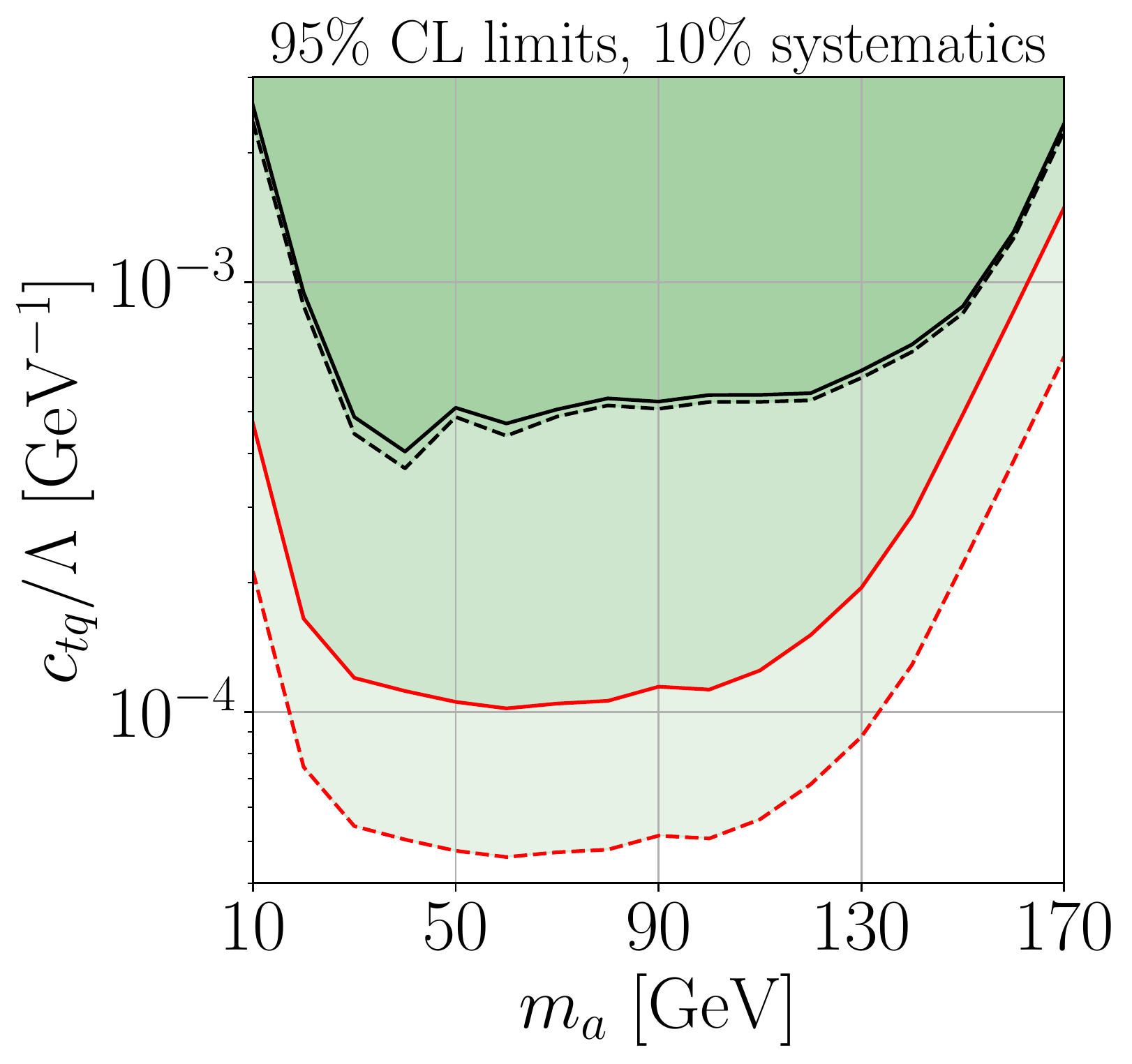}
        \includegraphics[width=0.45\linewidth]{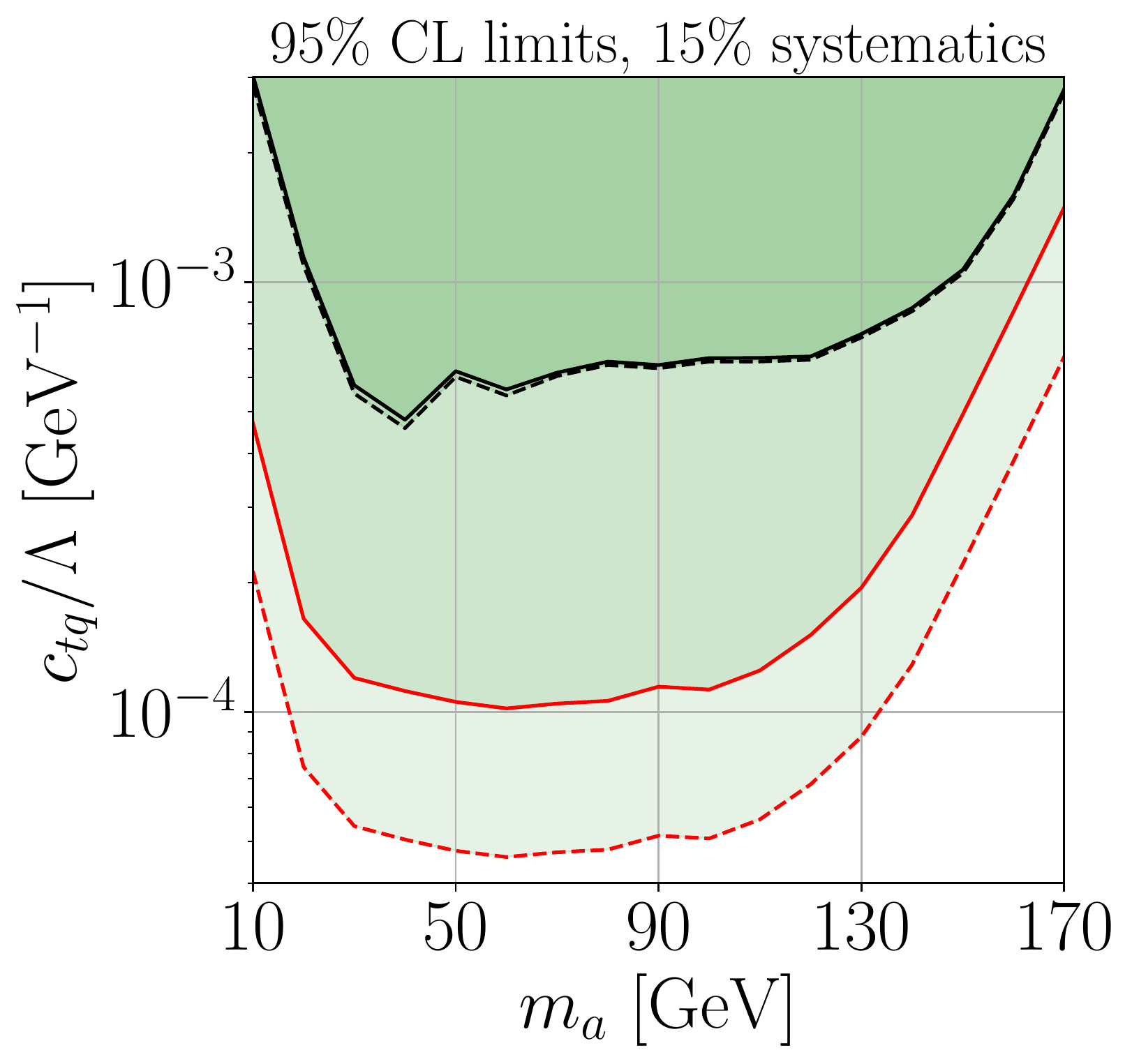}
        \caption{95\% CL exclusion region in the $m_a \times c_{tq}/ \Lambda$ parameter space. Red and black lines represent the $pp \to aa$ and $pp \to at$ processes, respectively. The solid and dashed lines represent the reach of these channels for $139 \, \si{fb^{-1}}$@13 TeV and $3000 \, \si{fb^{-1}}$@14 TeV limits, respectively. Each panel assumes a different level of systematic error. For systematic errors greater than $5\%$, the $pp \to aa$ process is the most competitive to exclude the model.}
        \label{fig:95cl}
    \end{figure}

Because the flavon decays almost exclusively into taus plus electrons or muons when $m_a < m_t$, our analysis is not sensitive to lepton couplings since these parameters cancel out when we add the branching ratios. Only the top-quark couplings $c_{tu}, c_{tc}$, and the flavon mass, $m_a$ can thus be probed in these channels. 

Figure \ref{fig:95cl} depicts the 95\% CL limits for both production channels. The confidence intervals take into account different levels of systematic errors. 
The red and black lines show the 95\% CL exclusion limits for the $pp \to aa$ and $pp \to at$ processes, respectively, and the shaded areas represent the excluded regions of the parameter space if no signal is observed. The solid and dashed lines represent the reach for integrated luminosities of $139 \, \si{fb^{-1}}$@13 TeV and $3000 \, \si{fb^{-1}}$@14 TeV, respectively. Double flavon production presents better prospects for exclusion in all systematic scenarios considered in this work. As shown in Figure \ref{fig:95cl}, the single production, which is contaminated by non-negligible backgrounds, is sensitive to systematic uncertainties. Yet, for flavon masses close to the top mass, single production wins over double production, complementing the exclusion limits. Moreover, the limit on $c_{tq}/\Lambda$ from double production, where the statistical significance is just $\sqrt{S}$, scale as $L^{-1/4}$ with the luminosity, and increasing data up to 3000 fb$^{-1}$ improves the sensitivity by a factor not much larger than 2.
\begin{figure}[ht]
        \centering
        \includegraphics[width=0.45\linewidth]{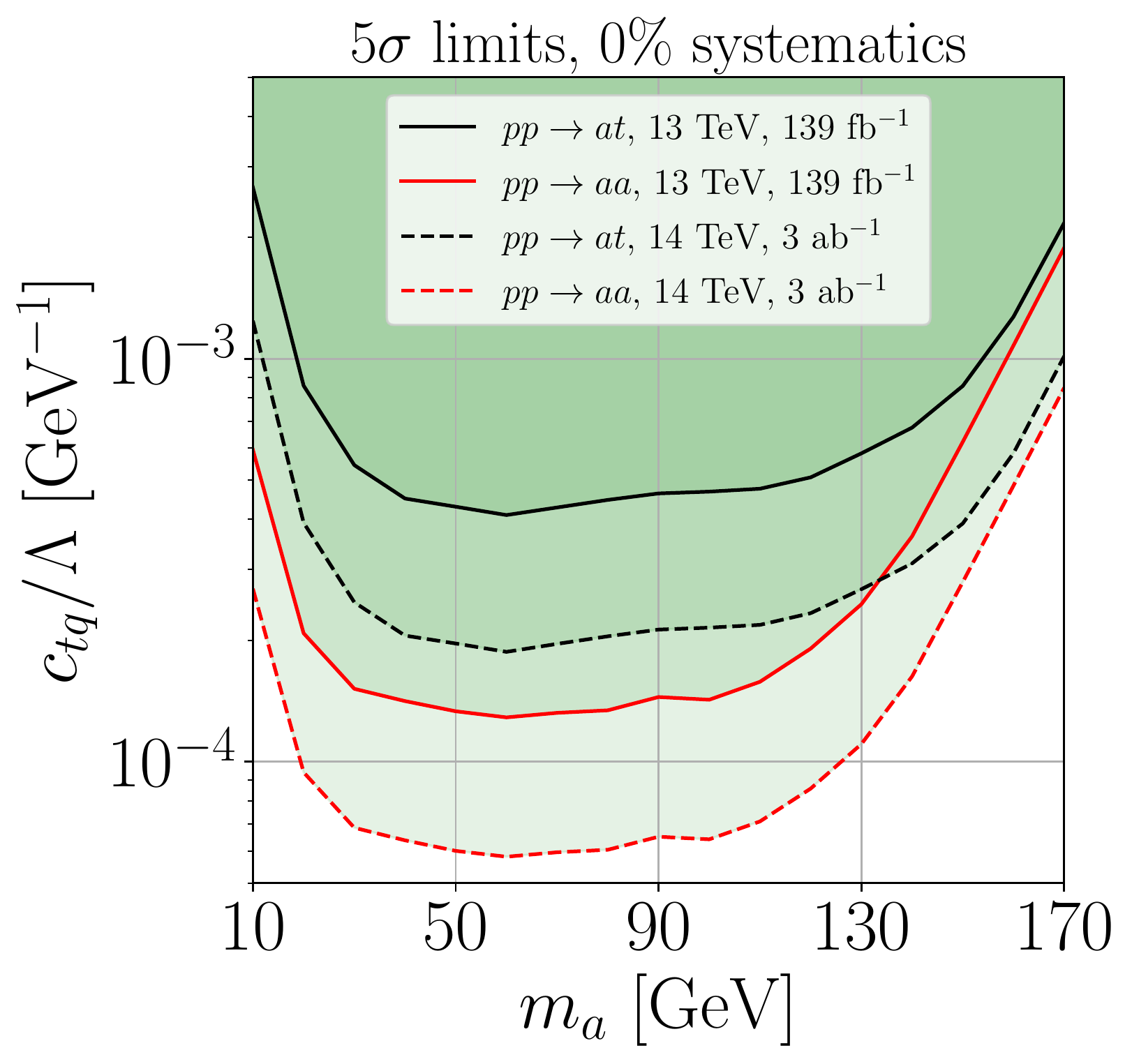}
        \includegraphics[width=0.45\linewidth]{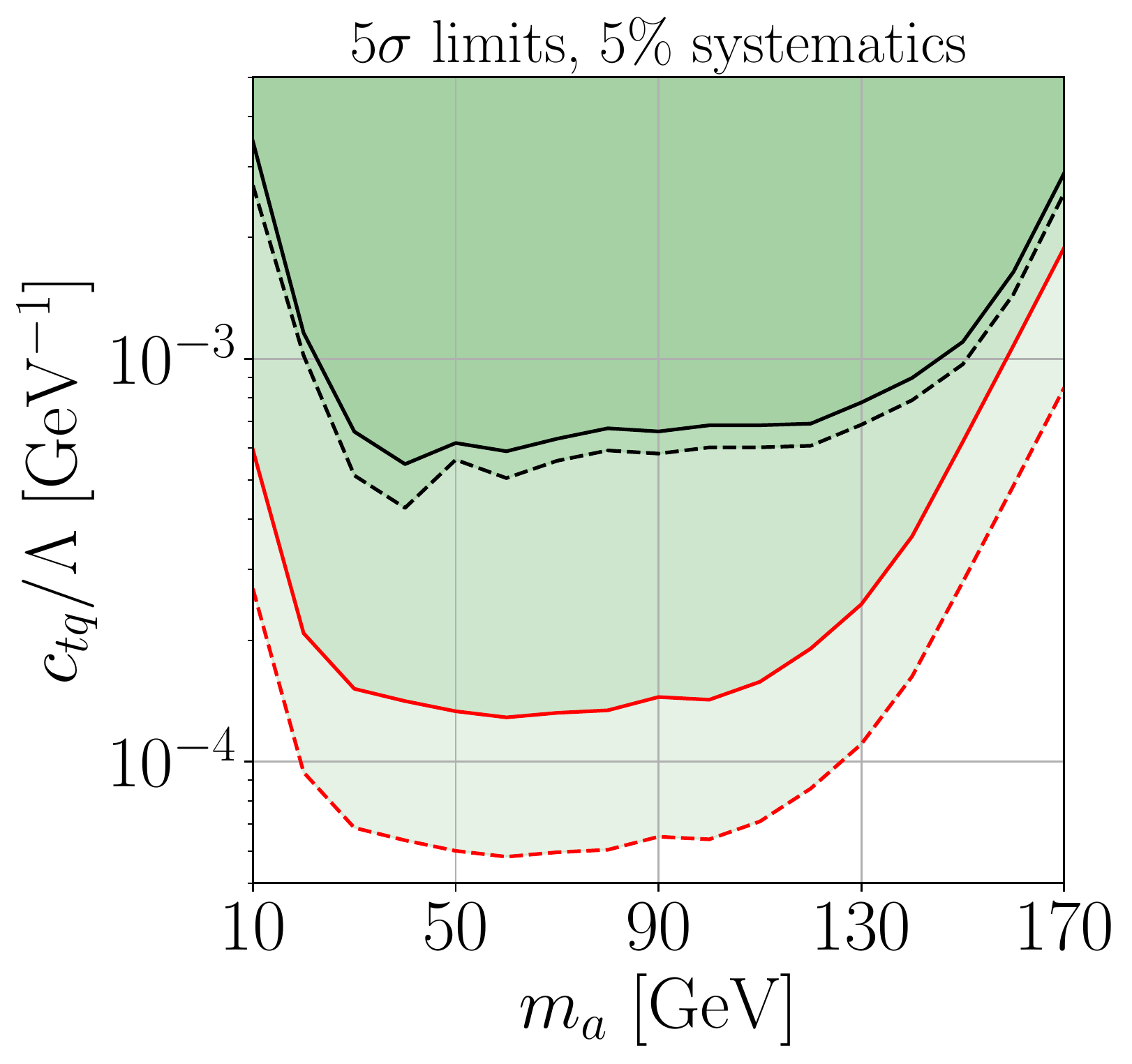}
        \includegraphics[width=0.45\linewidth]{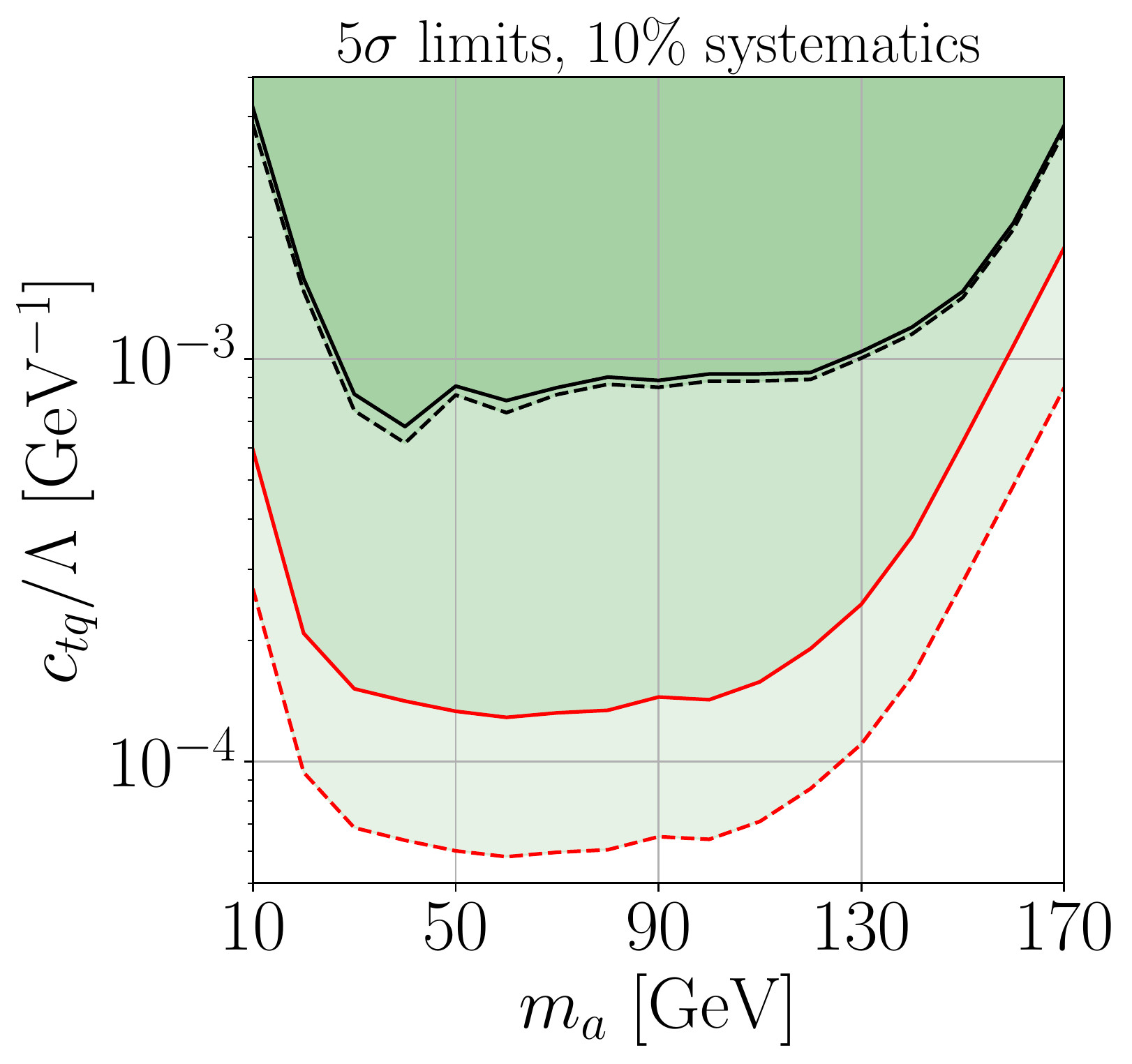}
        \includegraphics[width=0.45\linewidth]{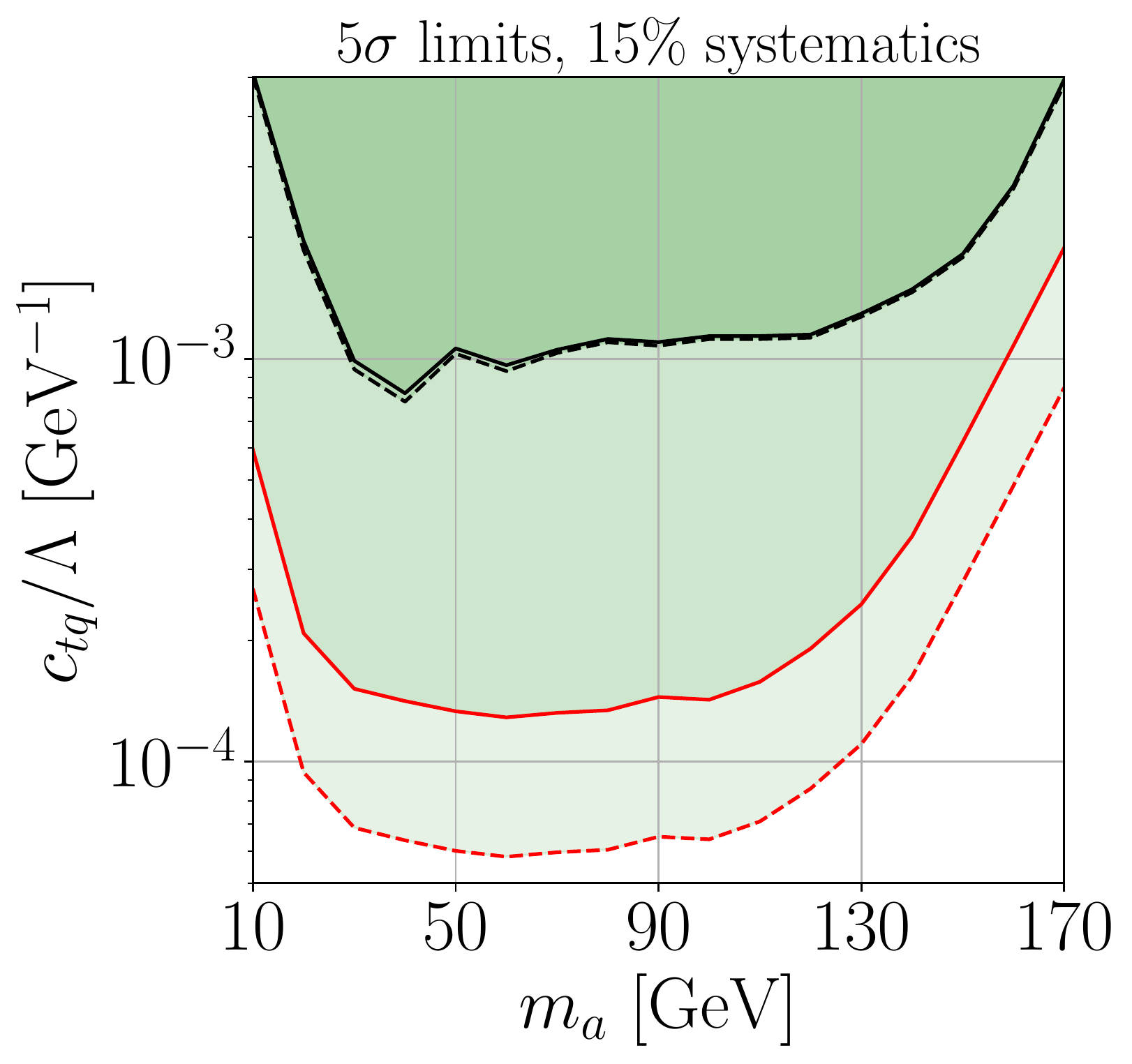}
        \caption{$5 \sigma$ discovery region in the $m_a \times c_{tq}/ \Lambda$ parameter space. Red and black lines represent the $pp \to aa$ and $pp \to at$ processes, respectively. The solid and dashed lines represent the reach of these channels for $139 \, \si{fb^{-1}}$@13 TeV and $3000 \, \si{fb^{-1}}$@14 TeV, respectively. Each panel assumes a different level of systematic error. For systematic errors in excess of $5\%$, the $pp \to aa$ process is the most competitive to discover a signal from this model.} 
        \label{fig:5sigma}
    \end{figure}

Fixing systematic errors on 5\%, effective coupling $c_{tq}/\Lambda$ of order $10^{-4}$($5\times 10^{-5}$) GeV$^{-1}$ can be excluded in double production for flavon masses around 80 GeV and integrated luminosity of 139 fb$^{-1}$@13 TeV (3000 fb$^{-1}$@14 TeV). Not observing single production events excludes $c_{tq}/\Lambda$ of order $10^{-3}$ for $150<m_a<170$ GeV irrespective of the luminosity.

The $5\sigma$ discovery regions for both channels are shown in Figure \ref{fig:5sigma}. We show the discovery potential for the current $139 \, \si{fb^{-1}}$@13 TeV (solid lines) and projected $3000 \, \si{fb^{-1}}$@14 TeV (dashed lines). Just like the 95\% CL limits, the discovery region from single flavon production depends on the systematic errors in the background predictions, and the single production becomes competitive $m_a$ closer to the top mass.

\subsection{Benchmarking model parameters}

Now that we established the region of $c_{tq}/\Lambda$ {\it versus} $m_a$ that can be probed at the LHC, we look for lepton-flavon couplings that escape theoretical and experimental constraints but keep compatibility with the current muon $(g-2)$ data. Our goal is to instantiate some model parameters to demonstrate their phenomenological viability. A thorough exploration of the parameter space might reveal other interesting spots, but it is beyond the scope of this analysis.


We want to find  LFV couplings that are compatible with the $(g-2)$ of the muon~\cite{Muong-2:2006rrc,Muong-2:2015xgu,Muong-2:2023cdq} 
data and lie inside the region of discovery of the LHC but evade the top decay constraint of Eq. (\ref{eq:top-decay}), the LFV-induced $\mu\to e\gamma$, $\tau\to e\gamma$, and $\tau\to \mu\gamma$  transitions of Eq. (\ref{eq:lfv-decay}), and respect the unitarity bounds of Eq. (\ref{eq:unitarity-bounds}).


The measured anomalous magnetic moment of the muon has been showing a longstanding discrepancy compared to the SM prediction of approximately $5\sigma$: $\Delta a_\mu = a_\mu^{\mathrm{exp}}-a_\mu^{\mathrm{SM}}=(24.9 \pm 4.9)\times 10^{-10}$, where $a_\mu\equiv (g-2)_\mu/2$. In the case of the electron, the situation seems inconclusive. With $a_e^{\mathrm{exp}}$ determined in \cite{Hanneke_2008,Hanneke:2010au}, the predicted value for $a_e^{\mathrm{SM}}$ 
using the fine-structure constant measured from $^{133}$Cs recoil leads to the discrepancy of $2.4\sigma$: $\Delta a_e = a_e^{\mathrm{exp}}-a_e^{\mathrm{SM}}=(-8.7\pm 3.6)\times 10^{-13}$~\cite{Parker:2018vye}. On the other hand, the value of $a_e^{\mathrm{SM}}$ obtained with the measured fine-structure constant measured using $^{87}$Rb furnishes a positive discrepancy of $1.6\sigma$: $\Delta a_e = a_e^{\mathrm{exp}}-a_e^{\mathrm{SM}}=(4.8\pm 3.0)\times 10^{-13}$~\cite{Morel:2020dww}. 
New measurements will be necessary to resolve the discrepancy, so we choose to work with the muon data only.



Since we are dealing with an EFT, some care needs to be taken with respect to the allowed range we vary the coupling constants, as it will affect the lowest consistent value of $\Lambda$, according to Eq. \eqref{eq:unitarity-bounds}. For example, if we focus on the parameter space region where $\Lambda > 500 \, \si{GeV}$, then we must respect $v_{e \mu}, a_{e \mu} \lesssim 2\times 10^4$, while all other four couplings must be smaller than approximately $10^3$. The higher we set the minimum $\Lambda$, the greater the intervals in which we can vary these constants. 

When flavons have masses much greater than the lepton masses, we can approximate the loop functions, shown in Appendix \ref{app:decay}, as follows:
\begin{equation}
    g_3(x_j) \approx \frac{2 \ln(x_j) - 3}{x_j} \,, \quad h_3(x_j) \approx \frac{2}{3x_j} \,, \quad g_4(x_j)\approx -\frac{1}{3x_j},
\end{equation}
where $x_j=(m_a/m_j)^2,\; j=e,\mu,\tau$.

The loop functions involved in $\Delta a_\mu$, $g_3(x_j)$ and $h_3(x_j)$, are always positive for $m_a>10$ GeV, which guarantees the muon anomaly will have the correct (positive) sign as long as we have $v_{\mu \tau} > a_{\mu \tau}$, accordingly to Eq. \eqref{eq:g-2}.

In general, the constraints from $\mu\to e+\gamma$, $\tau\to e+\gamma$, and $\tau\to \mu+\gamma$, depend on the couplings of the lepton inside the loop amplitude with the external leptons and the new physics scale, $g_{ij}g_{ik}/\Lambda^2$, where $i$ is the flavor of the lepton inside the loop and $m_j>m_k$ must be respected in order for the decay to be possible. This way, if we increase the couplings, the constraints extend to larger $\Lambda$, eventually excluding the points of the parameter space that are compatible with $(g-2)$ data and that can be discovered in the LHC.







\begin{figure}
    \centering
    \includegraphics[width=0.45\linewidth]{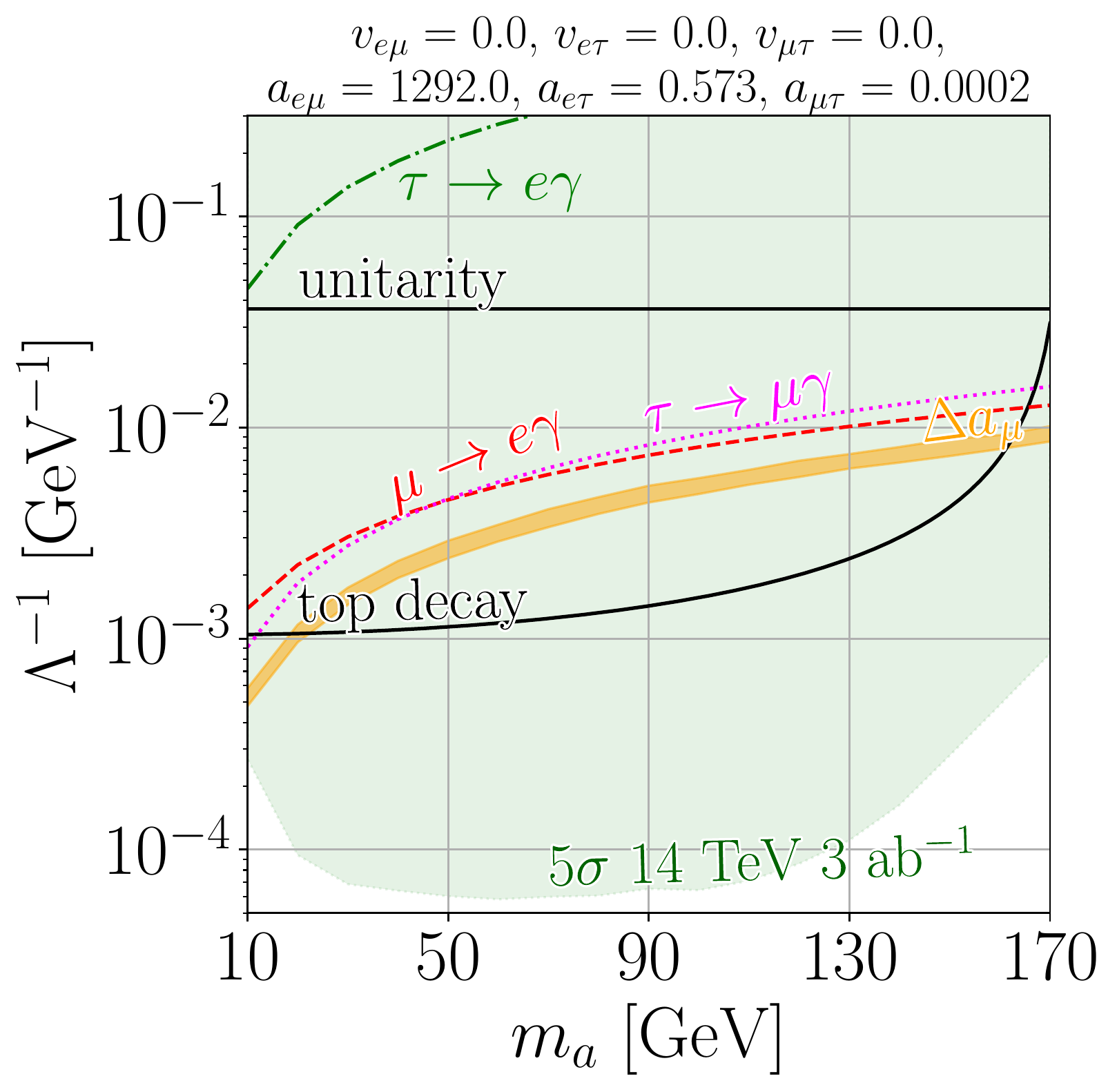}
    \includegraphics[width=0.45\linewidth]{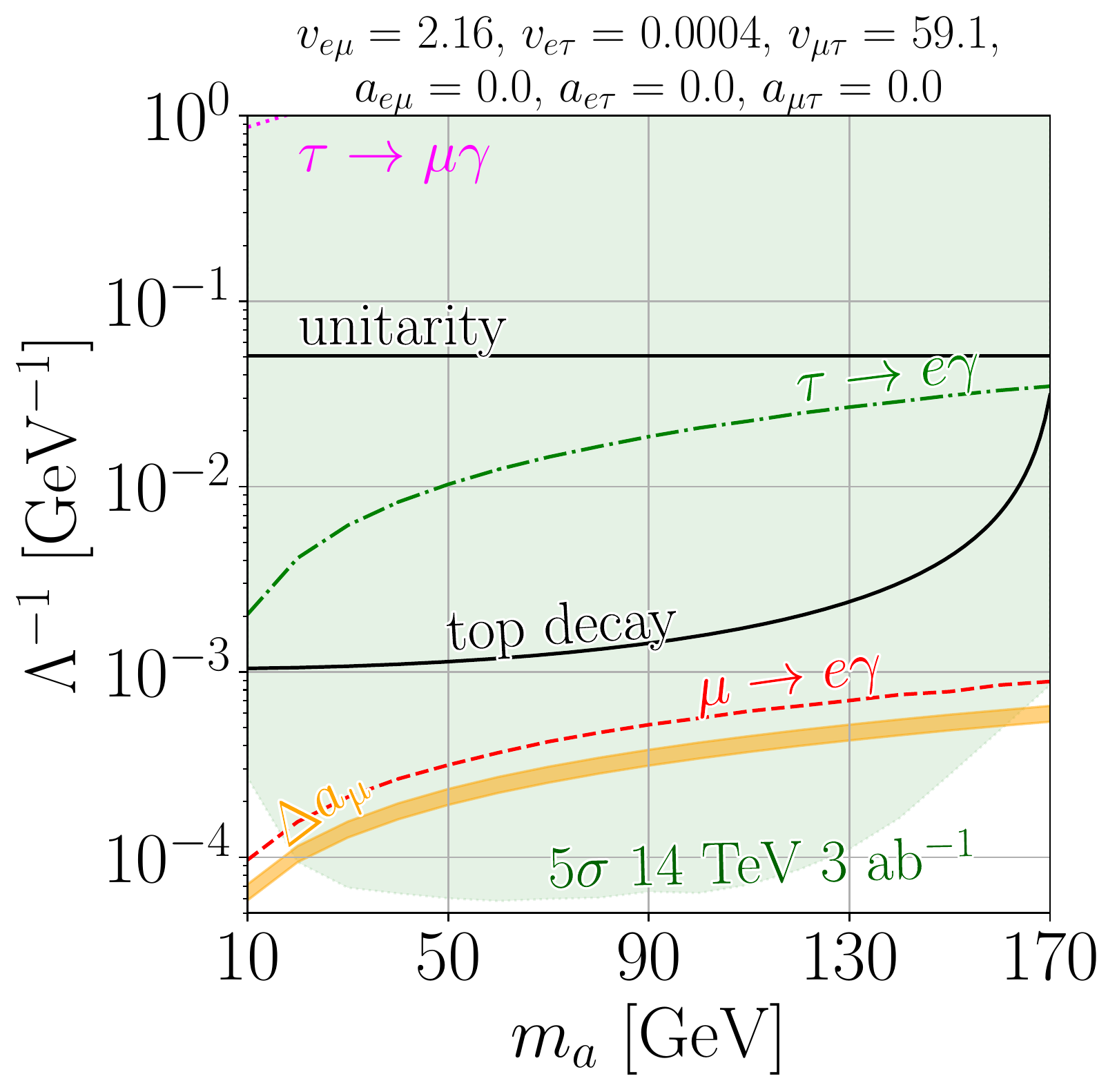}
    \includegraphics[width=0.45\linewidth]{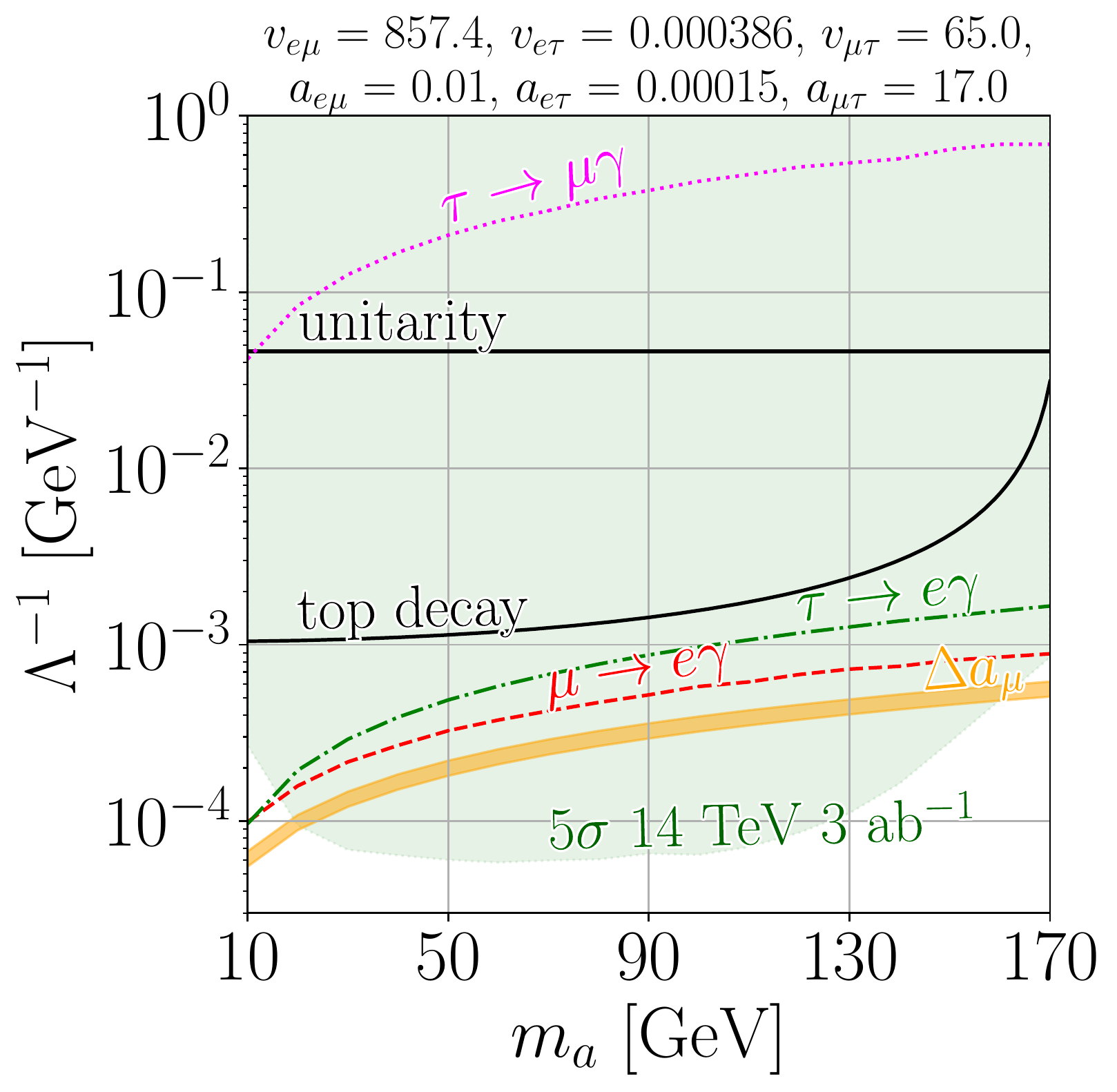}
    \includegraphics[width=0.45\linewidth]{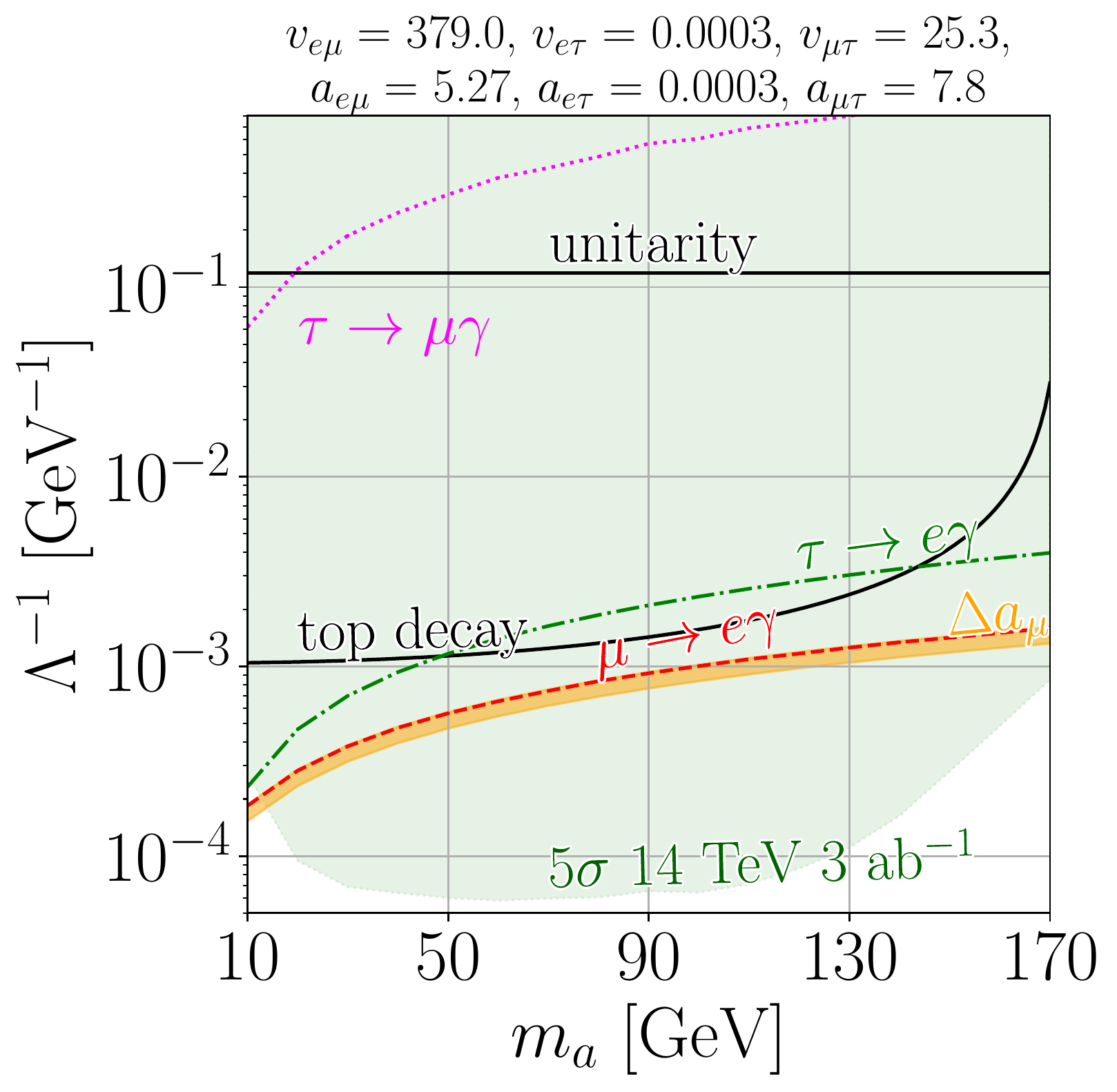}
    \caption{
    Examples of models with purely axial (upper left), purely vectorial (upper right), and all LFV couplings turned on (lower panels). The orange region can explain the muon anomaly within its $1 \sigma$ experimental uncertainty. In all panels, we found models that lie inside a region of the parameter space that can be probed at the LHC and evade all decay rates, the top quark, and unitarity constraints.
    }
    \label{fig:mixed}
\end{figure}


We performed a log-uniform scan over the $v_{e \mu}$, $a_{e \mu}$ couplings in the interval [$10^{-4}$, $2 \times 10^4$] and all other lepton couplings in the range $[10^{-4}, 1.3 \times10^3]$, looking for points in the LFV coupling space that evade all constraints from lepton decays, the top quark constraint explained in Section \ref{exp-const}, and the unitarity limits on $\Lambda$ described in Section \ref{sec:model}. After that, we perform a grid search in $\Lambda$ and $m_a$ to find a complete set of parameters that can explain the anomalous magnetic moment of the muon within its $1\sigma$ experimental band. The final requirement is lying in the region of the $c_{tq}/\Lambda$ {\it versus} $m_a$ plane that can be probed by the LHC.

With these requirements over the coupling constants, our random search found approximately 2.5\% of viable points for the general case of all LFV constants turned on. In the special case of all $v$ turned off, we found approximately 0.8\% of viable hyperspace points. When we turned off all $a$ instead, we found around 3.5\% viable solutions. For each of these three cases, the sample size was $10^5$.

In Figure \ref{fig:mixed}, we show some examples of successful models we found using the aforementioned method. In each panel, the orange band represents the region that accommodates the $(g-2)_\mu$ data. The upper left and upper right panels show special solutions for the purely axial and vectorial flavons, respectively. In the lower panels we depict two examples of general theories, where all couplings are turned on.

In the first panel of Figure \ref{fig:mixed}, we see that the orange band barely evades the top constraint. The reason lies in the setting of $v_{e \mu} = 0$, which is the constant with greater effect to push the $(g-2)_\mu$ to higher values of $\Lambda$. This is the case with the lowest percentage of solutions found, since $a_{\mu \tau} > v_{\mu \tau}$ and the correct sign of $\Delta a_\mu$ is not always guaranteed anymore, as can be seen from Eq. \eqref{eq:g-2}. In the second panel, we show a case of vectorial flavon where the $(g-2)$ band escapes all constraints easily.



The lower panels of Figure \ref{fig:mixed} represent a class of viable models that we identified where all LFV couplings are turned on. In both of those models, $a_{e \mu}$ is much larger than all other couplings. The reason for this can be traced back to the BR($\mu \to e \gamma$) derived from the decay rate in Eq. \eqref{eq:FandGemu} which is the most sensible constraint, but does not depend on $a_{e \mu}$ or $v_{e \mu}$. Thus, by increasing $a_{e \mu}$ we are able to make the $\Delta a_\mu$ band escape all constraints without changing the $\mu \to e \gamma$ line. Another feature present in the lower panels of Figure \ref{fig:mixed} is that the $a_{e \tau}$ and $v_{e \tau}$ are of order $10^{-4}$ and the $a_{\mu \tau}$ and $v_{\mu \tau}$ are of order 10. The reason is also related to BR($\mu \to e \gamma$) since its expression depends on all of these constants only. According to Eq. \eqref{eq:FandGemu}, this branching ratio excludes smaller $\Lambda$ if $v_{e \tau} a_{\mu \tau} \approx a_{e \tau} v_{\mu \tau}$ and $a_{e \tau} a_{\mu \tau} \approx v_{e \tau} v_{\mu \tau}$.

As a final benchmark point, we show a parameter space point with nonvanishing couplings that evade experimental and theoretical constraints, and explain $(g-2)$ data, to which both the LHC and \textcolor{blue}{the} MEG II experiment will present sensitivity: 
\begin{eqnarray}
\label{eq:lfv-consts-final}
&& a_{\mu e} = 400 \,, \quad a_{\tau e} = 1 \times 10^{-4} \,, \quad a_{\tau \mu} = 10\,, \nonumber \\
&& v_{\mu e} = 0.01 \,, \quad v_{\tau e} = 2 \times 10^{-4} \,, \quad v_{\tau \mu} = 40  \,.     
\end{eqnarray}

 \begin{figure}
     \centering
     \includegraphics[width=0.45\linewidth]{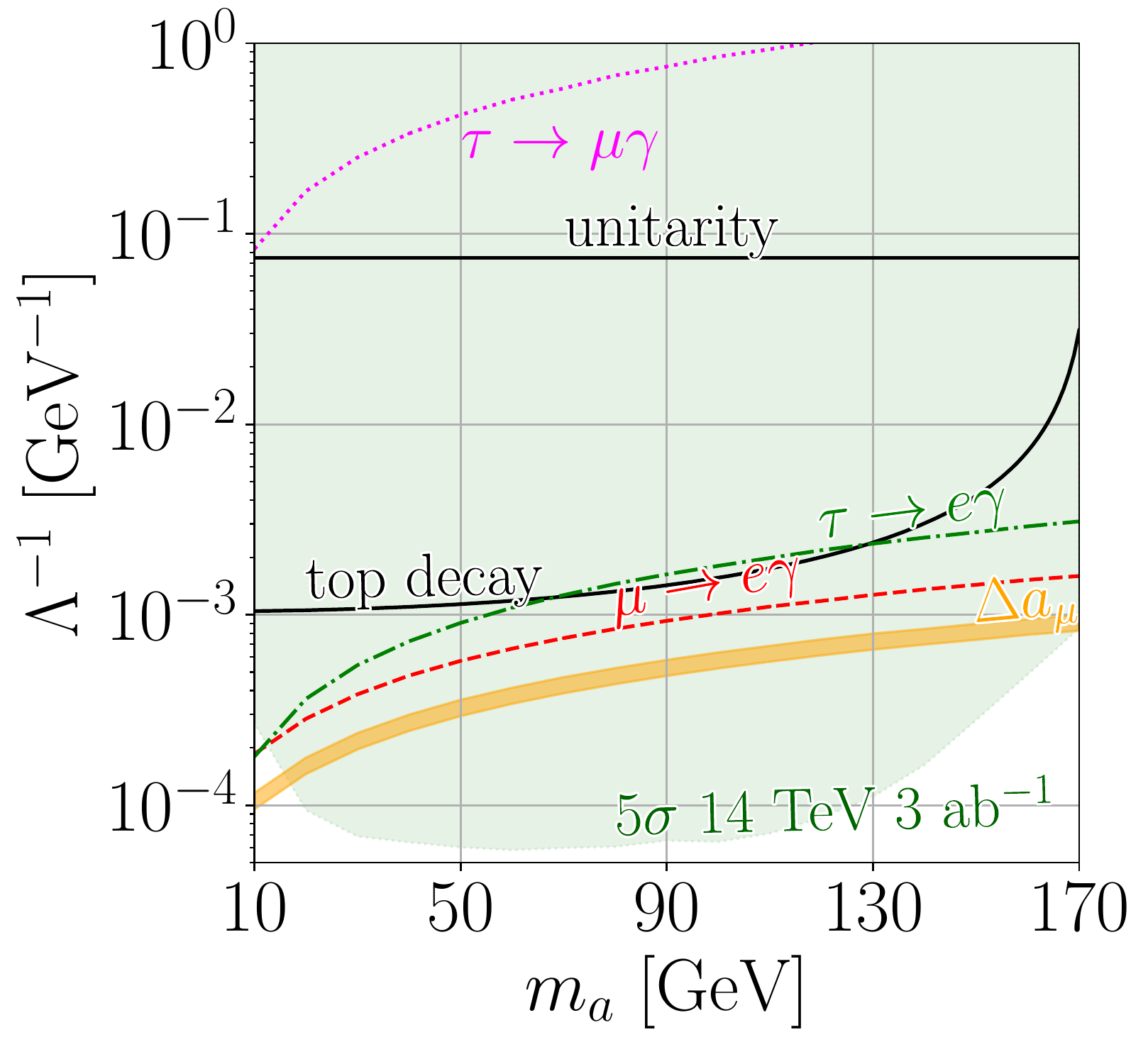}
     \includegraphics[width=0.45\linewidth]{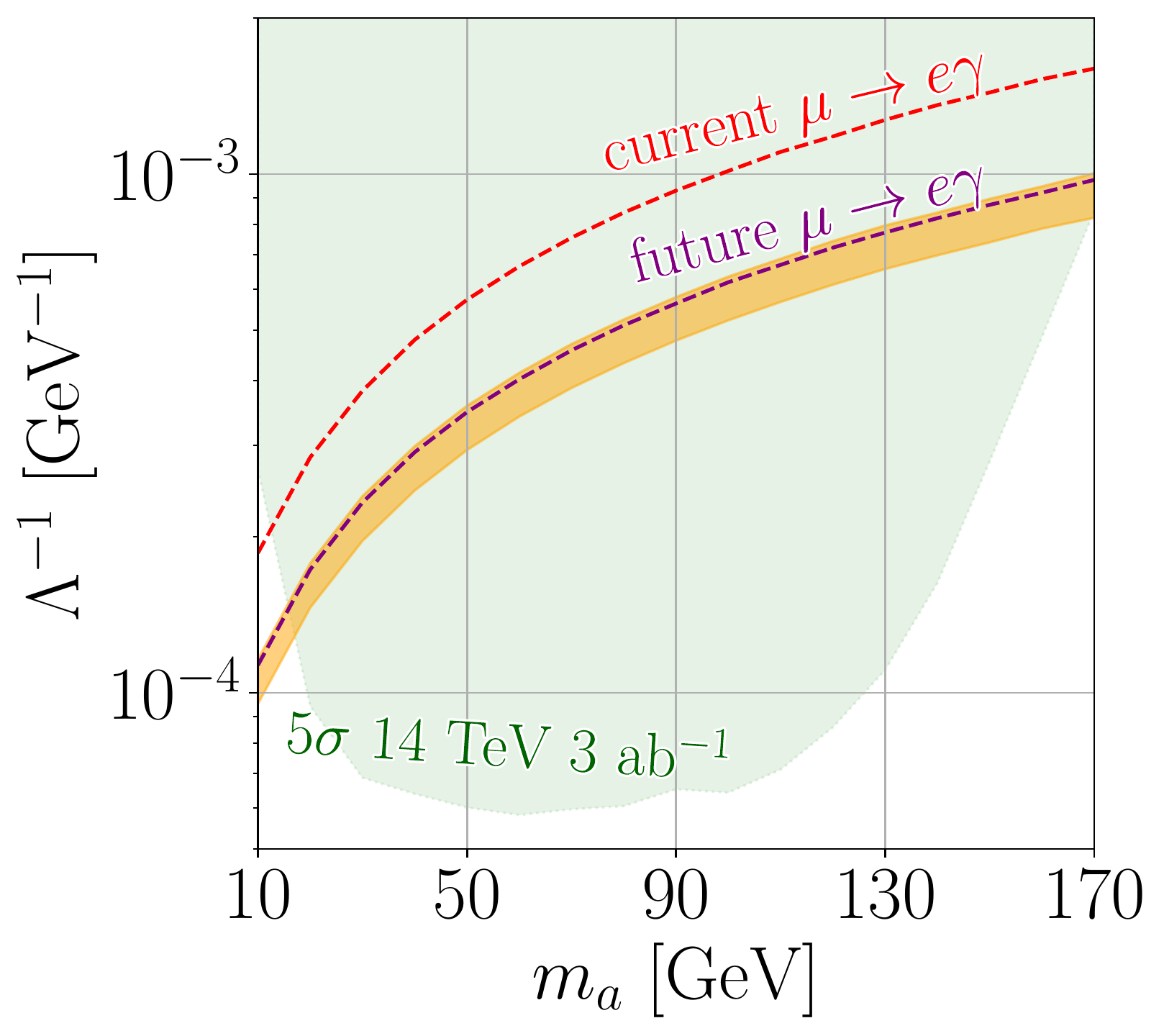}  
     \caption{Left: summarized results in our work for a specific set of LFV couplings described in Eq. (\ref{eq:lfv-consts-final}). The $(g-2)$ region 
     is shaded in orange, the LFV exclusion limits are the colored lines, and the top quark constraint and unitarity bound are the black lines. The green shade represents the $5 \sigma$ discovery threshold for the $pp \to aa$ process. Right: we plot the MEG-II future sensitivity for the  LFV couplings in Eq. (\ref{eq:lfv-consts-final}). For this specific set of couplings, we see that the $(g-2)$ region can be partially probed by the planned experiment.}
     \label{fig:g-2-new}
 \end{figure}

In the right panel of Figure~\ref{fig:g-2-new}, we forecast the constraints from the projected MEG II experiment and compare them with the current limits from MEG~\cite{MEG:2016leq}. As we see, part of the solutions of the scenario of Eq.~\eqref{eq:lfv-consts-final} can be probed in that future experiment. It is interesting to notice that both the LHC and MEG-II could work in a complementary way with respect to the detection of this model, since the preferred region, which explains the $(g-2)_\mu$ anomaly, is inside the reach of both experiments.

\subsection{Diagonal couplings}\label{sec:diag}

\begin{figure}
    \centering
    \includegraphics[width=0.45\linewidth]{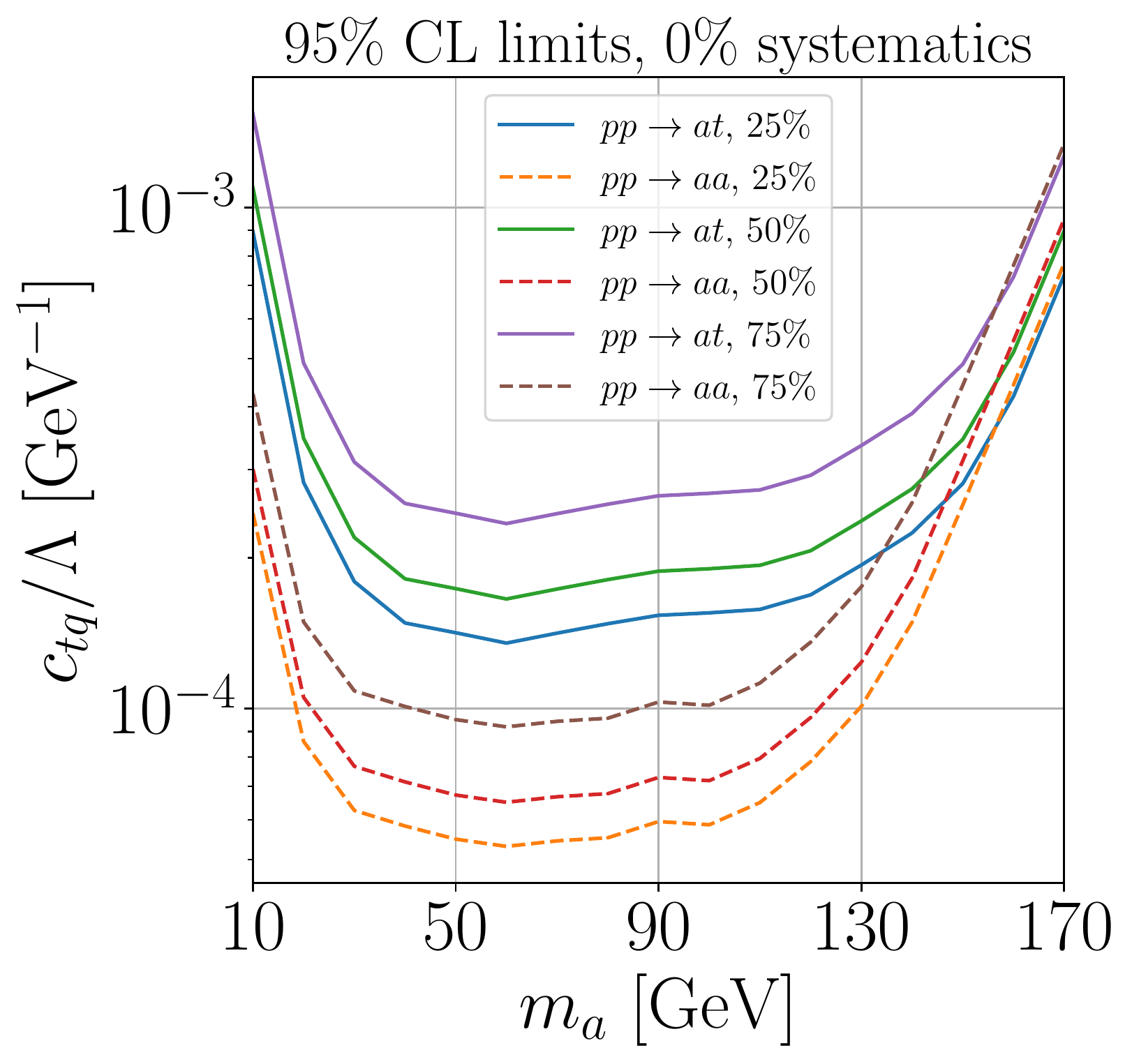}
    \includegraphics[width=0.45\linewidth]{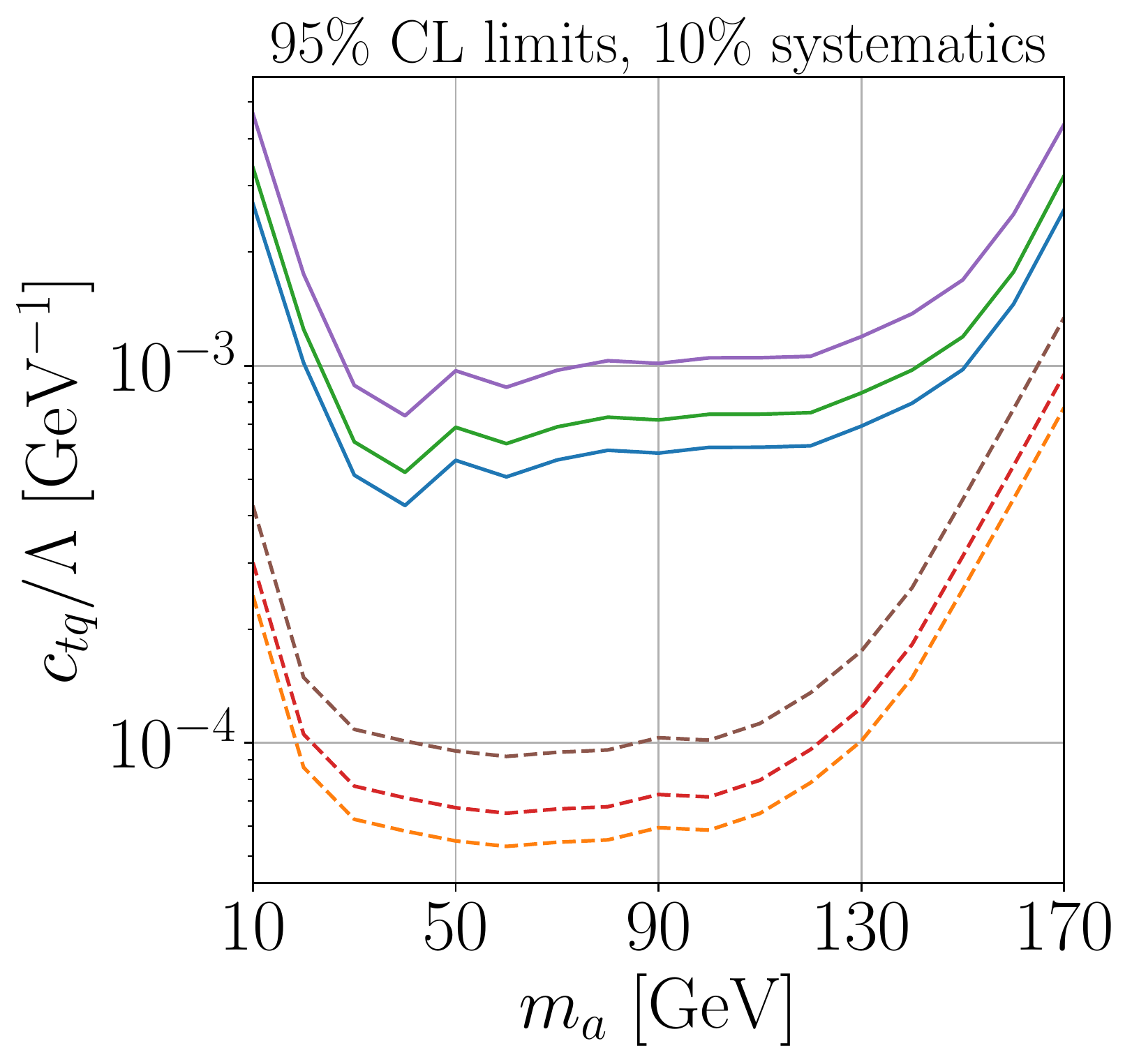}
    \includegraphics[width=0.45\linewidth]{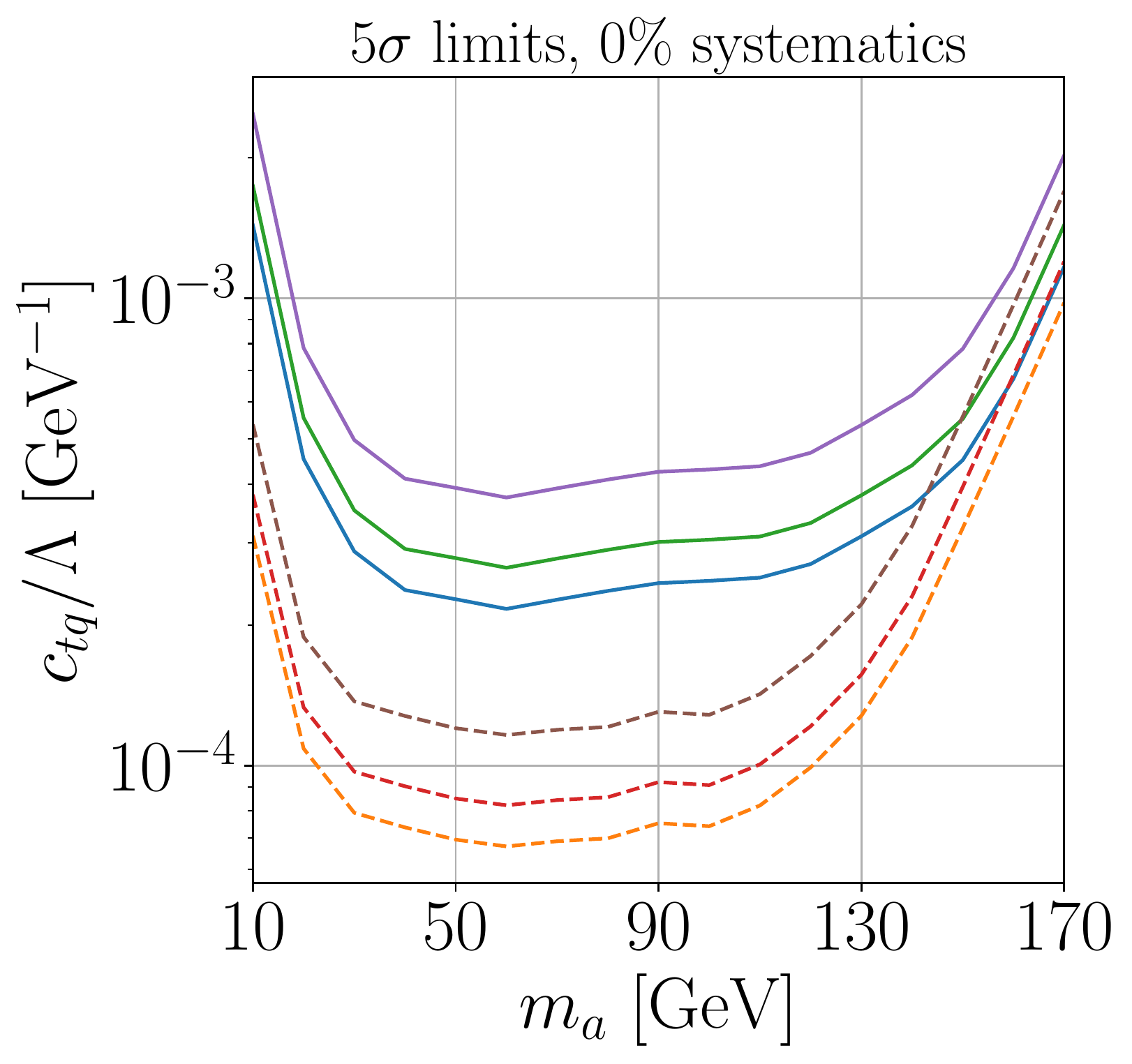}
    \includegraphics[width=0.45\linewidth]{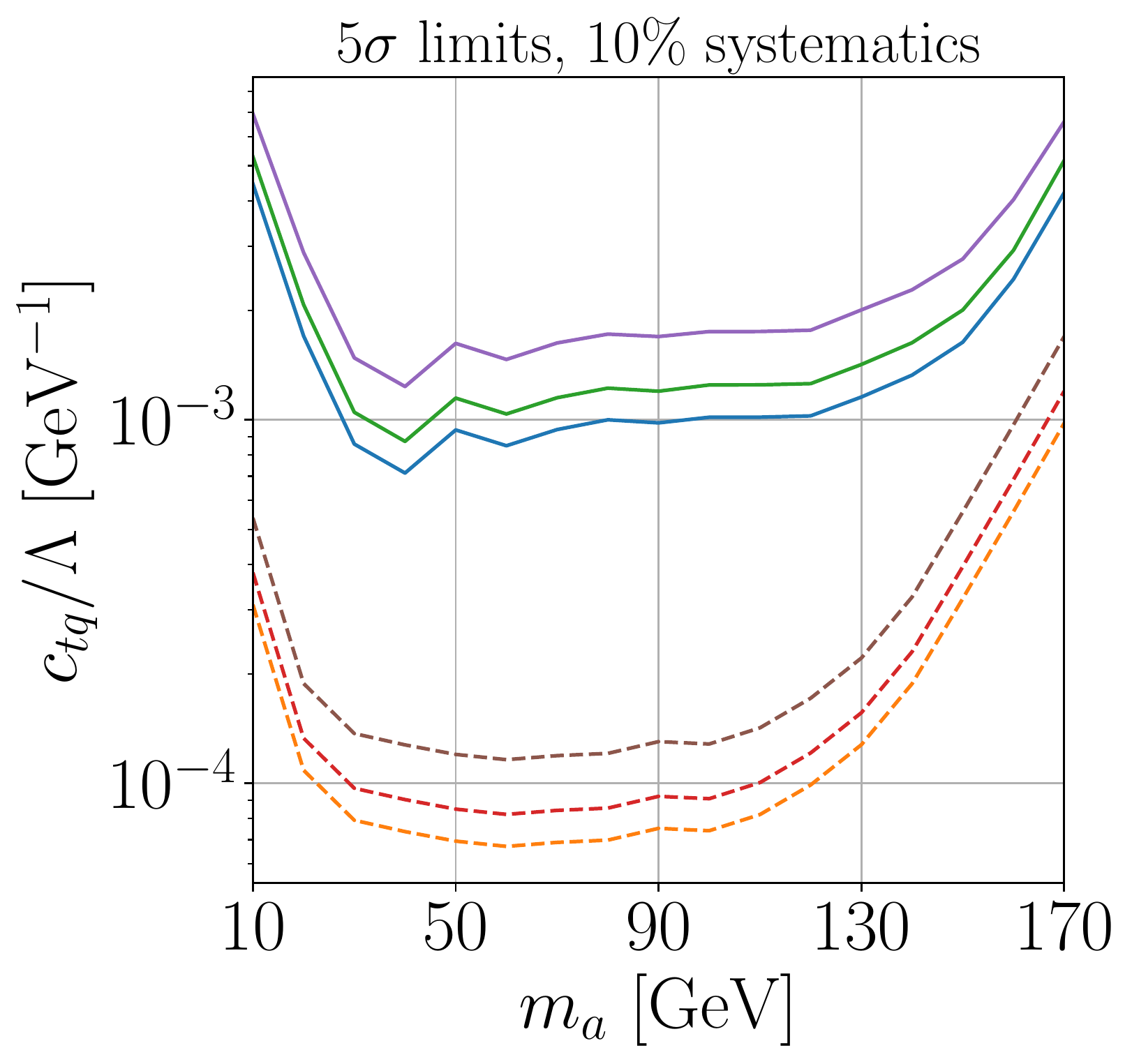}
    \caption{Effect of the diagonal couplings on the LHC exclusion (upper panels) and discovery (lower panels) regions for diagonal branching ratios of 25, 50, and 75\% for single and double production modes. We fix the integrated luminosity at 3000 fb$^{-1}$ and depict scenarios with (10\%) and without systematic uncertainties.}
    \label{fig:diagonal}
\end{figure}

The analysis presented so far focused on negligible diagonal couplings. However, most flavor theories should present sizeable diagonal couplings. As we are emphasizing, evading experimental constraints in the case where strong diagonal couplings leading to flavon decays into $gg,\gamma\gamma,WW,ZZ,Z\gamma,f_i\bar{f}_i$ might be difficult given the current status of searches for such scalars. Part of the motivation for this work is to investigate a testable EFT that evades such experimental constraints.

While evaluating the phenomenological viability of this extended scenario with the presence of diagonal couplings is beyond the scope of our analyses, we make the exercise of diluting the nondiagonal couplings to estimate the impact of diagonal couplings on the prospects of the LHC to exclude and discover our proposed signals. 

If $p$ represents the branching ratio of the flavon into all diagonal states, then the single production and decay of flavons is diluted by $1-p$, and the double production by $(1-p)^2$. The minimum effective flavon-top coupling that can be probed at some statistical confidence level scales equally for single and double production. The double production has virtually no background, so we estimate its statistical significance as the root of the number of signal events. Consequently, because the production cross section is proportional to the fourth power of the effective coupling, then $c_{tq}/\Lambda\sim (1-p)^{1/2}$. In the case of single production, $c_{tq}/\Lambda\sim (1-p)^{1/2}$ because the statistical significance is estimated roughly as $s/\sqrt{b}$ and the production cross section scales as the second power of the effective coupling.

We show, in Figure \ref{fig:diagonal}, the impact of diagonal decays with branching ratios of 25, 50, and 75\% in the LHC sensitivity to probe the $c_{tq}/\Lambda$ effective coupling as a function of the flavon mass in scenarios with 0 and 10\% systematic uncertainties in the background rates. Even in the case of a strong suppression of nondiagonal couplings by $0.25$, the LHC sensitivity after 3000 fb$^{-1}$ of data is around 0.1 TeV$^{-1}$ still with or without systematic uncertainties. This diminished impact of diagonal couplings is due to the lack of background events in the same-sign taus of the double production channel.

\begin{figure}[t]
    \centering
    \includegraphics[width=0.45\linewidth]{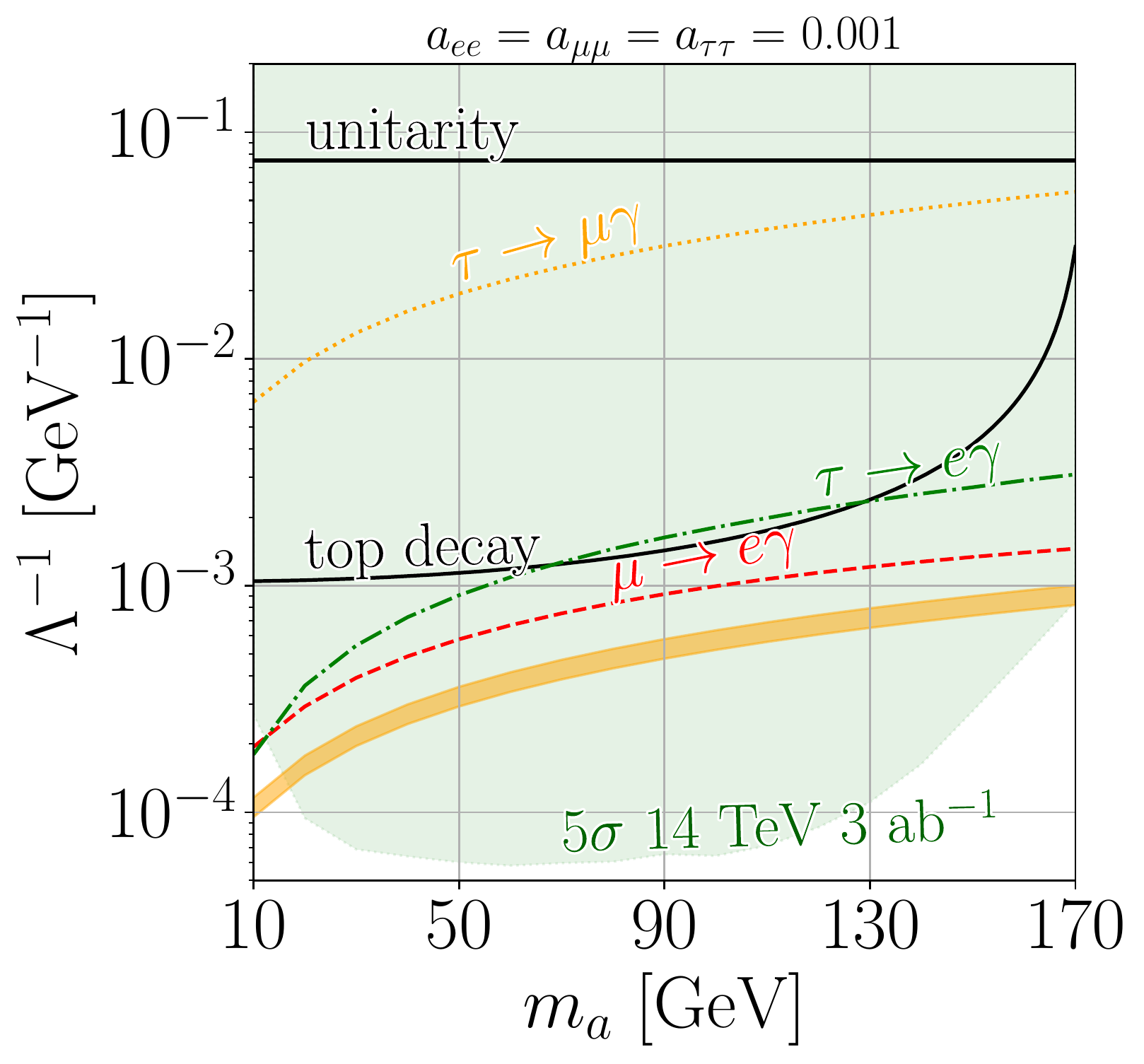}
    \includegraphics[width=0.45\linewidth]{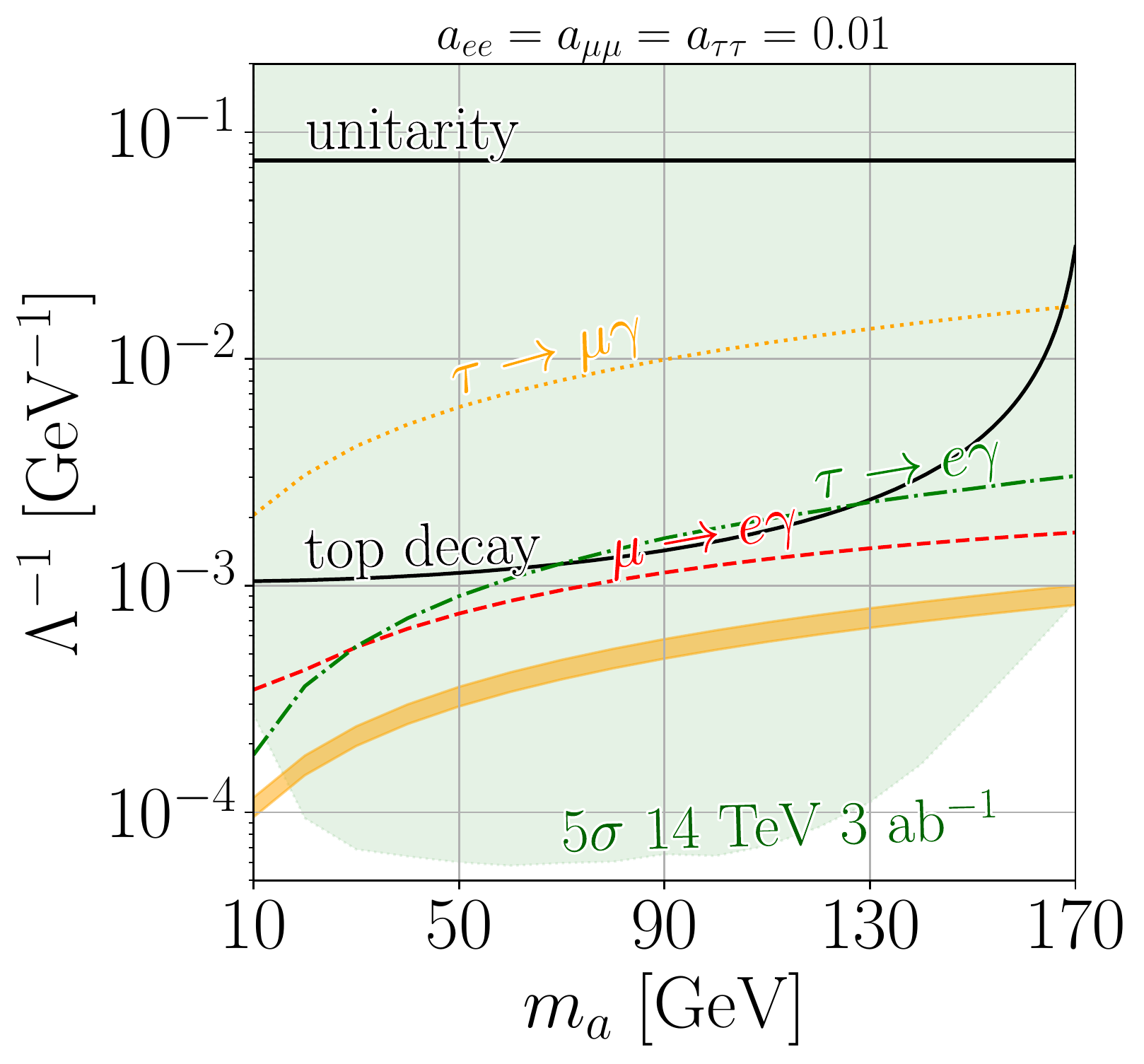}
    \includegraphics[width=0.45\linewidth]{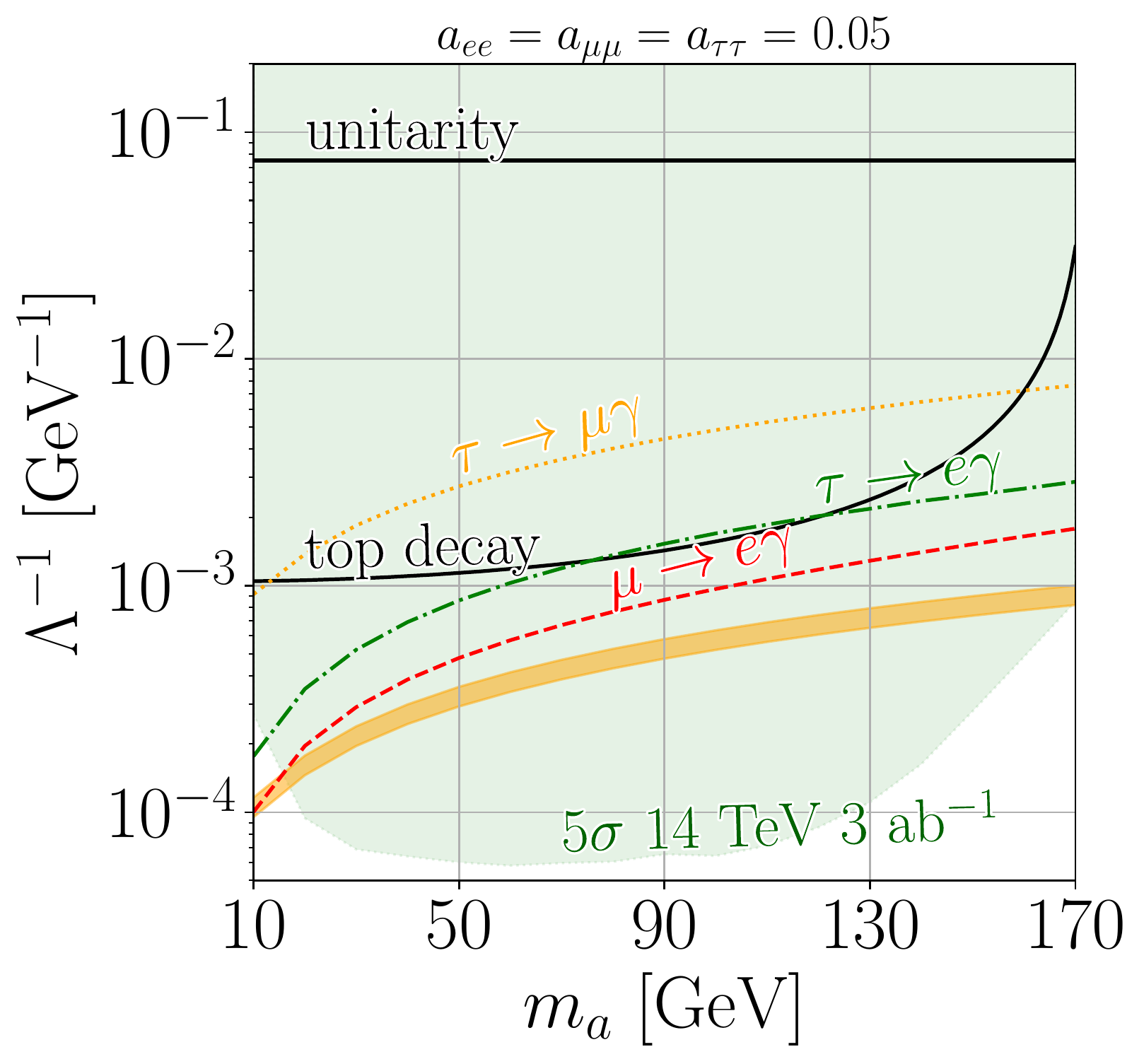}
    \includegraphics[width=0.45\linewidth]{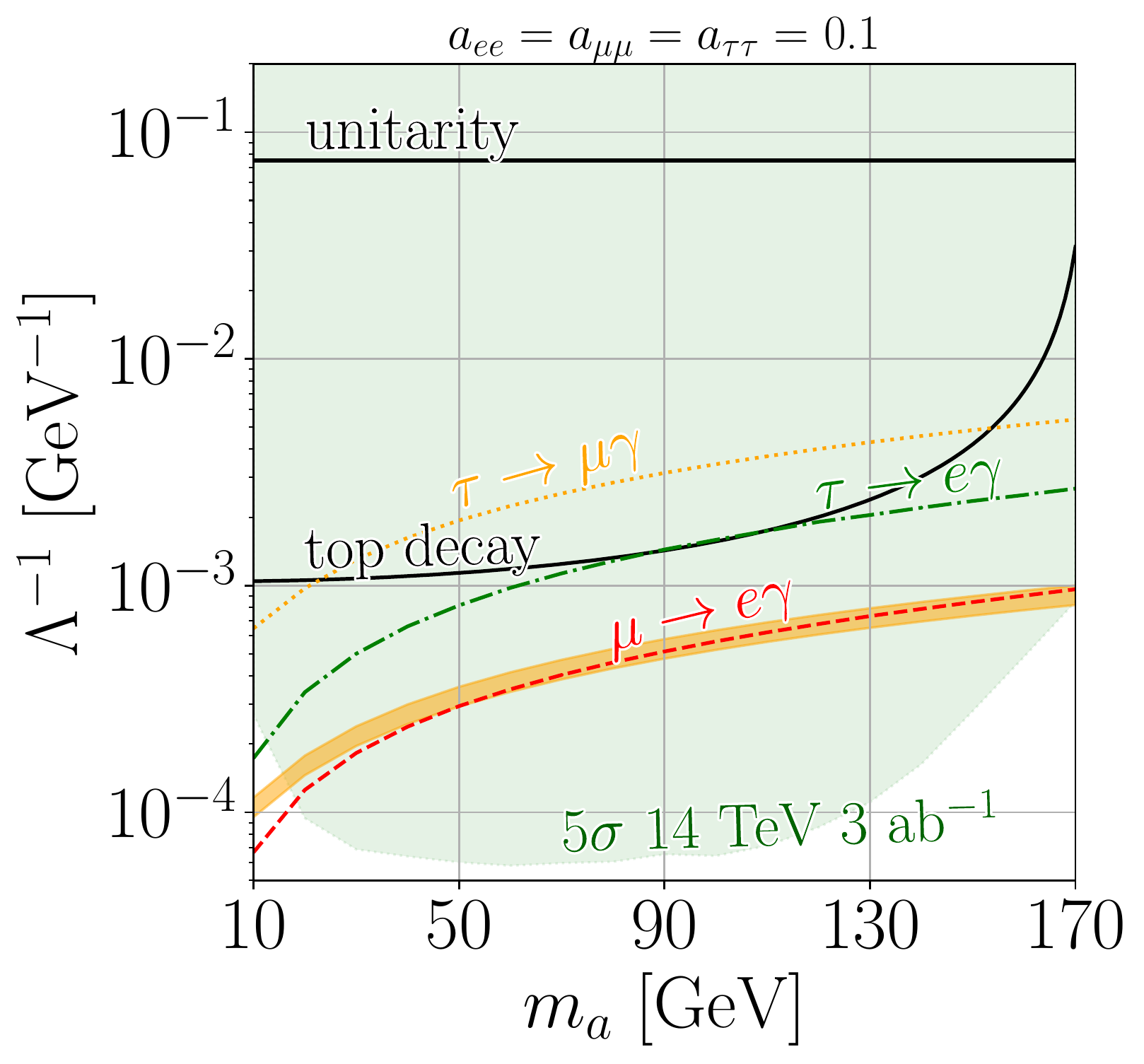}
    \caption{The effect of diagonal couplings on the lepton constraints and the region that accommodates the $(g-2)_{\mu}$ discrepancy. The values of the constants not shown in the figure are the same as in Eq. \eqref{eq:lfv-consts-final}. In each of these pictures, the $c_{a\gamma \gamma}\sim 10^{-8}$, thus highly suppressed. As we increase the diagonal couplings, it becomes harder to avoid all the constraints we considered in this work. For diagonal couplings of order 1, the model from Eq. \eqref{eq:lfv-consts-final} cannot explain the $g-2$ anomaly and evade the $\mu \rightarrow e \gamma$ bound simultaneously.}
    \label{fig:newg}
\end{figure}

In Figure \ref{fig:newg}, where nondiagonal couplings were fixed as in Eq.\eqref{eq:lfv-consts-final}, we show that as we increase the strength of the diagonal couplings, it becomes increasingly harder to evade all decay rate constraints and find model parameter instances that can be probed at the LHC. This behavior confirms our expectations that diagonal couplings make the model much more sensitive to experimental constraints and reinforces our motivation to seek maximally flavor violating scalars as a viable and attractive phenomenological model.

\section{Conclusions}\label{conclusao}

 Many extensions of the Standard Model can explain the discrepancy in the anomalous magnetic moment of muons and electrons. Among those extensions, new scalars that break leptonic and baryonic number conservation can be tested in hadron colliders like the LHC. Such flavor-violating scalars, or flavons, can also couple diagonally in flavor space. However, if they maximally violate flavor, coupling only nondiagonally in flavor space, various experimental constraints can be evaded. The challenge then consists in showing that these kinds of couplings can still explain the current experimental discrepancy in $(g-2)_\mu$, and also be tested in colliders.

 In this work, we demonstrated that maximally flavor-violating flavon models escape current experimental constraints, are compatible with the muon $(g-2)$ anomaly, and can be searched at the LHC with the current data. The High-Energy, High-Luminosity LHC will be capable of extending that search even further.

 In the models we tested, flavons couple to the mass of fermions, so it is easier to look for third-family signals. We concentrate our phenomenological analyses in the 10--170 GeV mass range which has not been worked out in the literature. The flavon decays almost exclusively into taus plus electrons/muons in this mass range. Because of the flavor violation, double scalar production might give rise to same-charge lepton signatures that present low SM background contamination. With default \texttt{Delphes3} $\tau$-jet tagging and supposing that the electric charge of hadronic taus can be efficiently measured after jet reconstructions, we found that double and associate flavon+top quark production channels can be used to probe the proposed flavon models. While associated $t+a$ production suffers from larger backgrounds and systematic uncertainties in the background rates, they can complement the double production for flavon masses close to the top quark mass.

 With 139 fb$^{-1}$, our results show that the 13 TeV LHC data already present sensitivity to those flavons. The proposed channel only probes the flavon couplings to top-up/charm quarks since the couplings to leptons cancel out in flavon branching ratios, almost 100\% into taus plus leptons. For flavon masses around 50--80 GeV, $c_{tq}/\Lambda$ of order 0.1/TeV can be excluded at 95\% CL. Discovery is also possible for larger couplings of 0.2/TeV for favorable cases but larger than 1/TeV for masses close to the top mass. Projecting the results for the HL-LHC with 14 TeV, those limits extend to 0.05/TeV for 95\% CL exclusion and 0.03/TeV for $5\sigma$ discovery in the most favorable cases. 

 We also found many instantiations of the models' parameters that explain the measured muon anomalous magnetic moment, evade LFV constraints, and can be reached by the LHC. Some general lessons could be learned from this exercise. First, flavons with purely vectorial couplings to fermions seem to be strongly disfavored being excluded by LFV constraints. Solutions with purely axial couplings are rare but not impossible in the range of parameters that we investigated. In particular, increasing $a_{\mu\tau}$ makes it easier to escape current experimental constraints but also pushes the solutions to $(g-2)$ to larger $\Lambda$, making it harder to test the models in the LHC.

 Scenarios with mixed vectorial-axial couplings give us the freedom to escape constraints, explain muon $(g-2)$ data, and present good prospects at the LHC. Moreover, the next generation of LFV experiments, like MEG II, will complement the search of the LHC for viable points of the maximally violating flavon model. 

Most UV complete models of flavor are expected to predict diagonal couplings. We showed that even if the flavon decays predominantly to diagonal modes, the double production of flavons and decay into same-sign taus with virtually no SM backgrounds is still a useful probe of the model at the LHC. However, increasing the importance of diagonal couplings makes it harder for the model to evade current experimental constraints and reinforces the motivation to look for maximally flavor-violating scalars and their interactions.
	
\bigskip{}
	
\textbf{Acknowledgments}: This study was financed in part by Conselho Nacional de Desenvolvimento Científico e Tecnológico (CNPq), via the Grants No. 307317/2021-8 (A. A.) and No. 305802/2019-4 (A. G. D.), and in part by the Coordenação de Aperfeiçoamento de Pessoal de Nível Superior – Brasil (CAPES) – Finance Code 001 (D. S. V. G.). A. A. also acknowledges support from the FAPESP (No. 2021/01089-1) Grant. E. d S. A. thanks FAPESP for its support  (Grant  No.  2022/07185-5). We also want to thank the anonymous referee for effective comments that improved the work.

\appendix

\section{Partial Wave Unitarity Bounds}
\label{appendix:unitarity}

In the two-to-two scattering the corresponding helicity amplitude can be expanded in partial waves in the center–of–mass system as \cite{JACOB2000774} 
\begin{equation}
 M_{fi}\left(f_{\lambda_1} f_{\lambda_2} \rightarrow f_{\lambda_3} f_{\lambda_4}\right) = 16\pi\, \sum_{J} (2\,J+1)D^{*J}_{\lambda_i\lambda_f}(R^{-1}(\varphi_i,\theta_i,0)R(\varphi_f,\theta_f,0))T^{J}_{\lambda_f,\lambda_i},
\end{equation}
where $\lambda$ refers to the helicity, $\lambda_i \,=\, \lambda_1 - \lambda_2$, $\lambda_f = \lambda_3 - \lambda_4$, $J$ the total angular momentum, $D^J_{\lambda_i \lambda_f}$ is the standard rotation matrix given by 
\begin{equation}
D^{s}_{m^{\prime}m}\left(R\left(\varphi\,,\,\theta\,,\,\gamma\right)\right) \,=\, \left\langle s\,,\,m'\right| U\left[R\left(\varphi\,,\,\theta\,,\,\gamma \right)\right] \,\left|s\,,\,m \right\rangle \,=\, e^{-i \,\varphi\, m'}\, d^{\,s}_{m^{\prime}m}\left(\theta\right)e^{-i\, \gamma\, m},
\end{equation}
$U$ is the unitary operator representing a rotation, $d^{s}_{m^{\prime}m}$ the Wigner rotation matrix, and $T^J_{\lambda_f\lambda_i} \,=\, \left\langle J,\,M,\,\lambda_3\,,\,\lambda_4|\,T(s)\,| J,\,M,\,\lambda_1\,,\,\lambda_2\right\rangle$. The partial-wave unitarity for the elastic channels requires that
\begin{equation}
\left|T^{J}_{\lambda_f \lambda_i}\left(f_{\lambda_1} f_{\lambda_2} \rightarrow f_{\lambda_3} f_{\lambda_4}\right) \right| \leq 1.  
\end{equation}
\indent Using the orthogonal relations of the $D^{J}_{\lambda_i \lambda_f}$ we get
\begin{equation}
T^J_{\lambda_f\lambda_i} \,=\, \frac{1}{32\pi}\, \int_{0}^{\,\pi} d\theta\, \sin\theta\, D^J_{\lambda_i\lambda_f}\left(0\,,\,\theta\,,\,0\right) M_{fi}.
\end{equation}
where we set the azimuthal angles and the initial polar angle to zero. \\
\indent More stringent bounds can be obtained by diagonalizing $T^J$ in the particle and helicity space and then applying the previous condition as in Ref. \cite{Lee}. To obtain this limit we consider the case where only the vector or axial coupling is different from zero for a specific flavor.
Since we have only one dimension-five operator, the possible combinations of states we have are 
$\left|f_i \bar{f}_j\right\rangle$, $\left|a f_i\right\rangle$ and $\left|a \bar{f}_i\right\rangle$, where $a$ is the flavon and $f_i$ is a fermion of generation $i$. The most stringent bound comes from the scattering of $f_i \bar{f}_j \rightarrow f_i \bar{f}_j$. The $T^J$ matrices for the vector and axial couplings in the basis $\left|f^i_+ \bar{f}^j_+\right\rangle$, $\left|f^i_- \bar{f}^j_-\right\rangle$, $\left|f^i_+ \bar{f}^j_-\right\rangle$, and $\left|f^i_- \bar{f}^j_+\right\rangle$ (Here +,- refers to the helicity of the particles) are given by
\begin{eqnarray}
T^J = -\frac{v_{ij}^2\left(m_i - m_j\right)^2}{16\pi\Lambda^2}
    \begin{pmatrix}
     1 & 1 & 0 & 0\\   
     1 & 1 & 0 & 0 \\
     0 & 0 & 0 & 0\\
     0 & 0 & 0 & 0
    \end{pmatrix}, \quad
T^J = -\frac{a_{ij}^2\left(m_i + m_j\right)^2}{16\pi\Lambda^2}
    \begin{pmatrix}
     -1 & 1 & 0 & 0\\   
     1 & -1 & 0 & 0 \\
     0 & 0 & 0 & 0\\
     0 & 0 & 0 & 0
    \end{pmatrix}
\end{eqnarray}

\indent Diagonalizing the $T^J$ matrix in both cases we can improve the limit since the bigger eigenvalue is 2 and -2 two for the vector and axial coupling; therefore the final result is
\begin{equation}
  \left|v_{ij} \right| < \sqrt{8\pi}\frac{\Lambda}{m_j\,-\,m_i} \;\;\mbox{and}\;\; \left|a_{ij} \right| <\sqrt{8\pi}\frac{\Lambda}{m_j\,+\,m_i}\, .
\end{equation}

\section{Flavon-mediated decays}
\label{app:decay}

In models with flavor-violating couplings mediated by a scalar particle $a$, the expression for the partial width of the top quark decaying into $q = u, c$ is given by 
\begin{equation}\label{eq:top-decay}
    \begin{split}
       \Gamma(t \to a q) &=  \frac{1}{16\pi m_t^3 \Lambda^2 } \sqrt{\Big(m_a^2 - (m_t + m_q)^2\Big) \Big(m_a^2 - (m_t - m_q)^2\Big)} \\
       &\times \Bigg[a_{tq}^2 (m_t+m_q)^2\Big((m_t - m_q)^2 - m_a^2 \Big)  + v_{tq}^2 (m_t - m_q)^2\Big((m_t + m_q)^2 - m_a^2\Big) \Bigg]
    \end{split}
\end{equation}
where $a_{tq}$, $v_{tq}$ are the axial and vector flavon-top-light quark couplings, $m_q$ is the light quark mass, and $m_t$ the top mass. The flavon mass is $m_a$ and the energy scale which defines the effective field theory approach, $\Lambda$.

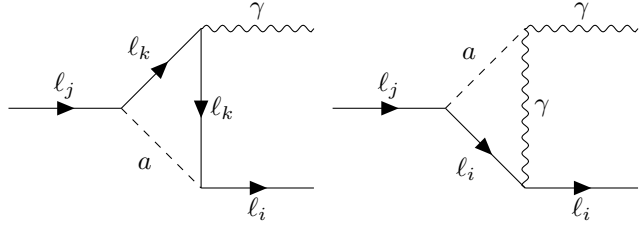
\begin{figure}
    \centering
    \begin{tikzpicture}[baseline=(current bounding box.center)]
    \begin{feynman}
        \vertex (b1);
        \vertex[right=of b1] (b2);
        \vertex[above right=of b2] (a3);
        \vertex[below right=of b2] (c3);
        \vertex[right=of a3] (a4);
        \vertex[right=of c3] (c4);
        \diagram* {
            (a3) -- [boson, edge label=\(\gamma\)] (a4),
            (b1) -- [fermion, edge label=\(\ell_j\)] (b2),
            (b2) -- [fermion, edge label=\(\ell_k\)] (a3),
            (b2) -- [scalar, edge label'=\(a\)] (c3),
            (a3) -- [fermion, edge label=\(\ell_k\)] (c3),
            (c3) -- [fermion, edge label'=\(\ell_i\)] (c4)
        };
    \end{feynman}
\end{tikzpicture}
    \begin{tikzpicture}[baseline=(current bounding box.center)]
    \begin{feynman}
        \vertex (b1);
        \vertex[right=of b1] (b2);
        \vertex[above right=of b2] (a3);
        \vertex[below right=of b2] (c3);
        \vertex[right=of a3] (a4);
        \vertex[right=of c3] (c4);
        \diagram* {
            (a3) -- [boson, edge label=\(\gamma\)] (a4),
            (b1) -- [fermion, edge label=\(\ell_j\)] (b2),
            (b2) -- [fermion, edge label'=\(\ell_i\)] (c3),
            (b2) -- [scalar, edge label=\(a\)] (a3),
            (a3) -- [boson, edge label=\(\gamma\)] (c3),
            (c3) -- [fermion, edge label'=\(\ell_i\)] (c4)
        };
    \end{feynman}
\end{tikzpicture}
    \caption{Lepton decay $\ell_j \to \ell_i \gamma$. When the model is maximally flavor violating, only the first diagram shows up, with $\ell_k \neq \ell_j, \ell_i$. When the diagonal couplings are turned on, not only is the second diagram also present, but in the first diagram it is allowed to have $\ell_k = \ell_j, \ell_i$.}
    \label{fig:lep-decay}
\end{figure}

Turning our attention to the leptons, the decay width for a heavier lepton $\ell_j$ into a lighter one $\ell_i$ plus an on shell photon is given by \cite{Cornella:2019uxs}
\begin{equation}\label{eq:lfv-decay}
    \Gamma (\ell_j \to \ell_i \gamma) = \frac{m_{\ell_j}}{8 \pi} e^2 \left( \abs{\mathcal{F}_{ij}(0)}^2 + \abs{\mathcal{G}_{ij}(0)}^2 \right) \,,
\end{equation}
where $m_j$ is the mass of the heavier lepton, $m_\mu$ or $m_\tau$. The Feynman diagram of these decays is depicted in Figure \ref{fig:lep-decay}. In the above equation, $\mathcal{F}_{ij}$ and $\mathcal{G}_{ij}$ have three contributions each. The linear
\begin{equation}
    \begin{split}
        \mathcal{F}^{ij}_2(0)_{\mathrm{lin}} &= -e^2 \frac{m_{\ell_j}^2}{8 \pi^2 \Lambda^2} a_{ij} c_{a \gamma \gamma} g_{\gamma}(x_j) \,, \\
        \mathcal{G}^{ij}_2(0)_{\mathrm{lin}} &= -e^2 \frac{m_{\ell_j}^2}{8 \pi^2 \Lambda^2} v_{ij} c_{a \gamma \gamma} g_{\gamma}(x_j) \,,
    \end{split}
\end{equation}
and the quadratic
\begin{equation}
    \begin{split}
        \mathcal{F}^{ij}_2(0)_{\mathrm{quad}} &= -\frac{m_{\ell_j}}{16 \pi^2 \Lambda^2} a_{ij} \Big[ a_{jj} m_{\ell_j} g_1(x_j) + a_{ii} m_{\ell_i} g_2(x_j)   \Big] \,, \\
        \mathcal{G}^{ij}_2(0)_{\mathrm{quad}} &= -\frac{m_{\ell_j}}{16 \pi^2 \Lambda^2} v_{ij} \Big[ a_{jj} m_{\ell_j} g_1(x_j) - a_{ii} m_{\ell_i} g_2(x_j)   \Big] \,.
    \end{split}
\end{equation}
For the $\mu \to e \gamma$ decay, we also have
\begin{equation}\label{eq:FandGemu}
    \begin{split}
        \mathcal{F}^{\mu e}_2(0)_{\text{quad}} &= -\frac{m_\mu m_\tau}{32 \pi^2 \Lambda^2} (a_{\tau \mu} a_{\tau e}  - v_{\tau \mu} v_{\tau e} ) g_3(x_\tau) \\
        \mathcal{G}^{\mu e}_2(0)_{\text{quad}} &= -\frac{m_\mu m_\tau}{32 \pi^2 \Lambda^2} (a_{\tau \mu} v_{\tau e}  - v_{\tau \mu} a_{\tau e}) g_3(x_\tau) \,.
    \end{split}
\end{equation}
where $a_{ji}$, $v_{ji}$ are the axial and vectorial LFV coupling constants connecting two leptons, $\ell_j$ and $\ell_i$, with the flavon, respectively. The variable $x_j$, $j=\mu, \tau$ is defined as
\begin{equation}
    x_j = \frac{m_a^2}{m_j^2}-i\eta,\; \eta\to 0^+ \,.
\end{equation}

For the $\tau \to \mu \gamma$ case, we have
\begin{equation}\label{eq:FandGeta}
    \begin{split}
        \mathcal{F}^{\tau \mu}_2(0)_{\text{quad}} &= -\frac{m_\mu m_\tau}{32 \pi^2 \Lambda^2} (a_{\tau e} a_{\mu e} + v_{\tau e} v_{\mu e}) g_4(x_\tau) \\
        \mathcal{G}^{\tau \mu}_2(0)_{\text{quad}} &= +\frac{m_\mu m_\tau}{32 \pi^2 \Lambda^2} (a_{\tau e} v_{\mu e} + v_{\tau e} a_{\mu e}
        ) g_4(x_\tau) \,.
    \end{split}
\end{equation}
There is also the $\tau \to e \gamma$ decay, where
\begin{equation}\label{eq:FandGmuta}
    \begin{split}
        \mathcal{F}^{\tau e}_2(0)_{\text{quad}} &= -\frac{m_\mu m_\tau}{32 \pi^2 \Lambda^2} (a_{\tau \mu} a_{\mu e} - v_{\tau \mu} v_{\mu e}) g_4(x_\tau) \\
        \mathcal{G}^{\tau e}_2(0)_{\text{quad}} &= -\frac{m_\mu m_\tau}{32 \pi^2 \Lambda^2} (a_{\tau \mu} v_{\mu e} - v_{\tau \mu} a_{\mu e}) g_4(x_\tau) \,.
    \end{split}
\end{equation}

\begin{figure}
    \centering
    \begin{tikzpicture}[baseline=(current bounding box.center)]
    \begin{feynman}
        \vertex (c1) {\(\ell_j\)};
        \vertex[right=of c1] (c2) ;
        \vertex[right=of c2] (c3);
        \vertex[right=of c3] (c4) {\(\ell_j\)};
        \vertex[above right=1.06cm of c2] (b);
        \vertex[below right=1.06cm of b] (x);
        \vertex[above=of b] (a) {\(\gamma\)};
        \diagram* {
            (c1) -- [fermion] (c2);
            (c2) -- [scalar, edge label'=\(a\)] (c3);
            (c3) -- [fermion] (c4);
            (c2) -- [fermion, edge label=\(\ell_k\)] (b);
            (b) -- [fermion, edge label=\(\ell_k\)] (x);
            (a) -- [boson] (b)
        };
    \end{feynman}
\end{tikzpicture}
    \caption{Feynman diagram of the main contribution to $(g-2)$ from lepton flavor-violating couplings.}
    \label{fig:g-2-tikz}
\end{figure}
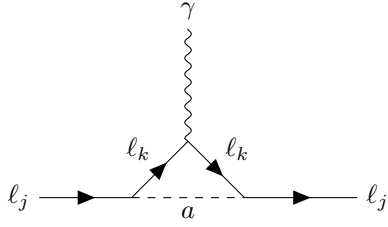

The $(g-2)_{\mu}$ uncertainty limits are codified by our parameters through the expressions
\begin{equation}
\label{eq:g-2}
        (\Delta a_\mu)_{\rm{LFV}} = \frac{m_\mu^2}{16 \pi^2 \Lambda^2} \left[ \qty(\abs{v_{e\mu}}^2+ \abs{a_{e\mu}}^2)h_3(x_\mu) + \frac{m_\tau}{m_\mu} \qty( \abs{v_{\mu \tau}}^2 - \abs{a_{\mu \tau}}^2)g_3(x_\tau)  \right] \,,
\end{equation}
for the lepton-flavor violating contributions, and
\begin{equation}
    \begin{split}
        (\Delta a_{\ell_i})_{\rm{LFC}} &=  -\frac{m_{\ell_i}^2}{16 \pi^2 \Lambda^2} \left[ 64\pi \alpha_{\mathrm{em}} c_{a \gamma \gamma} a_{ii} \left(\ln \frac{\Lambda^2}{m_{\ell_i}^2} - h_2(x_i) \right) + 4 \abs{a_{ii}}^2 h_1(x_i)  \right]
    \end{split}
\end{equation}
for the lepton-flavor conserving ones. The diagram for the main flavor-conserving contribution is depicted in Figure~\ref{fig:g-2-tikz}.

The loop functions are
\begin{equation}
    \begin{split}
        g_\gamma(x) &= 2 \ln \frac{\Lambda^2}{m_a^2} - \frac{\ln x}{x - 1} - (x-1) \ln \frac{x}{x-1} - 2 \,,\\
        g_1(x) &= \frac{x-3}{x-1}x^2 \ln x + 1 -2x - 2 x^{3/2} \sqrt{x-4} \ln \left( \frac{\sqrt{x}+ \sqrt{x-4}}{2} \right) \,, \\
        g_2(x) &= 1 - 2x + 2(x-1)x \ln \frac{x}{x-1} \,, \\
        g_3(x) &= \frac{2x^2 \ln x - (x-1)(3x-1)}{(x-1)^3} \,, \\
        g_4(x) &= 1 - 2x + 2(x-1)x \ln \frac{x}{x-1}  \,, \\
        h_1(x) &= 1 + 2x - (x-1)x \ln x + 2x(x-3)\sqrt{\frac{x}{x-4}} \ln \left( \frac{\sqrt{x}+ \sqrt{x-4}}{2} \right) \,, \\
        h_2(x) &= 1 + \frac{x^2}{6} \ln x - \frac{x}{3} - \frac{x+2}{x}\sqrt{(x-4)x} \ln \left( \frac{\sqrt{x}+ \sqrt{x-4}}{2} \right) \,, \\
        h_3(x) &= 2x^2 \ln \frac{x}{x-1} -1 -2x\,.
    \end{split}
\end{equation}

\section{Loop-induced decays to bottom quarks}
\label{app:loop}
\begin{figure}
    \centering
    \begin{tikzpicture}[baseline=(current bounding box.center)]
    \begin{feynman}
        \vertex (b1);
        \vertex[right=of b1] (b2);
        \vertex[above right=of b2] (a3);
        \vertex[below right=of b2] (c3);
        \vertex[right=of a3] (a4);
        \vertex[right=of c3] (c4);
        \diagram* {
            (a3) -- [fermion, edge label=\(b\)] (a4),
            (b1) -- [scalar, edge label=\(a\)] (b2),
            (b2) -- [fermion, edge label=\(t\)] (a3),
            (b2) -- [anti fermion, edge label'=\(\overline{u}\)] (c3),
            (a3) -- [boson, edge label=\(W\)] (c3),
            (c3) -- [anti fermion, edge label'=\(\overline{d}\)] (c4)
        };
    \end{feynman}
\end{tikzpicture}
\begin{tikzpicture}[baseline=(current bounding box.center)]
    \begin{feynman}
        \vertex (b1);
        \vertex[right=of b1] (b2);
        \vertex[above right=of b2] (a3);
        \vertex[below right=of b2] (c3);
        \vertex[right=of a3] (a4);
        \vertex[right=of c3] (c4);
        \diagram* {
            (a3) -- [fermion, edge label=\(b\)] (a4),
            (b1) -- [scalar, edge label=\(a\)] (b2),
            (b2) -- [fermion, edge label=\(t\)] (a3),
            (b2) -- [anti fermion, edge label'=\(\overline{c}\)] (c3),
            (a3) -- [boson, edge label=\(W\)] (c3),
            (c3) -- [anti fermion, edge label'=\(\overline{s}\)] (c4)
        };
    \end{feynman}
\end{tikzpicture}
    \caption{Feynman diagrams of the loop-induced contribution to $a\to bq$ decays.}
    \label{fig:loop-tikz}
\end{figure}
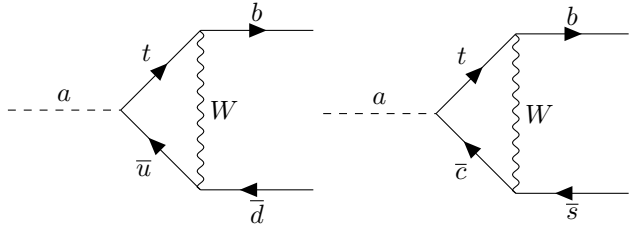

 Even in the absence of sizeable tree-level couplings between flavons and bottom quarks, FCNC currents with light quarks can be generated at the loop level as depicted in Figure~\ref{fig:loop-tikz}. In this case, strong couplings with the top quark mass could lead to $a\to bq,\; q=\bar{s},\bar{d}$ and diminish the branching ratio of the flavon into tau leptons. Not only this, but it also could lead to strong constraints on those quark couplings from the lack of evidence of FCNC interactions. 

 Neglecting light quark masses, including the bottom quark mass, in the numerator of the loop amplitude whenever it is suppressed by a top, $W$, or flavon mass, but keeping the mass dependence in the propagators, we can cast the loop amplitude into the form
 \begin{equation}
     {\cal M} = i\left(\frac{e}{\sqrt{2} s_W}\right)^2\left(\frac{mt}{\Lambda}\right) V_{tb} V_{cs} (v_{tq}-a_{tq})\frac{m_t}{m_W}\times \frac{1}{16\pi^2}\times \overline{u}_b(p_1)(F+G\gamma_5)v_s(p_2)
 \end{equation}
 where $F$ and $G$ and form factors from the loop amplitude are given right below. Here, $s_W$ is the sine of the Weinberg angle, and $e$ is the electron charge. The spinor corresponding to the final state bottom is $\bar{u}_b(p_1)$, while $v_s(p_2)$ is the antistrange spinor. The momentum of the scalar $a$ is $p_1+p_2$, where $(p_1+p_2)^2=m_a^2=m_b^2+m_s^2+2p_1\cdot p_2$.

 We see that in the limit where the flavon coupling to top quarks is of the $V+A$ type, $v_{tq}\to a_{tq}$, the coupling to $W$ bosons vanishes. 

 Assuming $V_{tb}=V_{cs}=1$ and taking into account all the four decay possibilities involving a bottom quark with a strange and a down quark, the partial decay width is given by
 \begin{equation}
     \Gamma_{bq} = \frac{3\alpha_{em}^2}{64\pi^3 s_W^4}\left(\frac{m_t^2}{m_W \Lambda}\right)^2 (v_{tq}-a_{tq})^2(F^2+G^2) m_a
 \end{equation}

 Taking $m_a=100$ GeV, $\Lambda=1$ TeV, $v_{tq}=1, a_{tq}=0$, $\alpha_{em}=1/137$, and $s_W^2=0.23$, this partial width is of order $10^{-8}$ GeV. The unitarity constraint of Eq.~\eqref{eq:unitarity-bounds} applied to the top quark bounds $v_{tq}(a_{tq})$ to less than $\sim 30$, for $\Lambda=1$ TeV, which is 1 to 2 orders of magnitude smaller than the partial width to tau leptons. The results were evaluated at the renormalization scale $\mu=m_W$, but the results do not change more than an order of magnitude even for light quark mass settings:

 \begin{eqnarray*}
F &=& -\frac{2 m_b^2+m_W^2}{4
   m_b m_W} -\frac{m_a^2 \left(m_b^2+m_W^2\right)+2 m_b^2
   \left(m_t^2+2 m_W^2\right)}{4 m_a^2 m_b
   m_W} B(m_b^2, mc, m_W) \\
   & + & \frac{m_b (m_t^2+2 m_W^2)}{2 m_a^2
   m_W} B(m_a^2 + m_b^2 + m_s^2, m_c, m_t)+\frac{m_b \left(m_t^2+2 m_W^2\right)
   \left(m_a^2+m_b^2+m_t^2\right)}{4 m_a^2 m_W
   \left(m_a^2+m_b^2\right)}\log(m_c^2/m_t^2) \\
   &-& \frac{m_a^2 \left(m_b^4+m_W^4\right)+2 m_b^2
   \left(m_b^2+m_W^2\right) \left(m_t^2+2
   m_W^2\right)}{8 m_a^2 m_b^3 m_W}\log(m_c^2/m_W^2) \\
   &+& \frac{m_s^4-m_s^2 \left(2
   m_t^2+m_W^2\right)+m_t^4+m_t^2 m_W^2-2
   m_W^4}{4 m_s m_W m_a^2}\log(m_t^2/m_W^2) -\frac{m_b}{4 m_W}\left(\frac{1}{\bar{\varepsilon}}+\log(\mu^2/m_c^2)\right)\\
   &-& \frac{m_b \left(m_t^4+m_t^2 m_W^2-2
   m_W^4\right)}{2 m_a^2 m_W} C_0(0, 0, m_a^2, 0, m_W, m_t) \\
   &+& \frac{m_s \left(m_s^2-m_t^2-2 m_W^2\right)}{2
   m_W m_a^2} B(m_s^2, m_t, m_W) \\
   G &=& \frac{2 m_b^2+m_W^2}{4 m_b
   m_W} +\frac{m_a^2 \left(m_b^2+m_W^2\right)+2 m_b^2
   \left(m_t^2+2 m_W^2\right)}{4 m_a^2 m_b
   m_W} B(m_b^2, mc, m_W) \\
   &-& \frac{m_b \left(m_t^2+2 m_W^2\right)}{2 m_a^2
   m_W} B(m_a^2 + m_b^2 + m_s^2, m_c, m_t)-\frac{m_b \left(m_t^2+2 m_W^2\right)
   \left(m_a^2+m_b^2+m_t^2\right)}{4 m_a^2 m_W
   \left(m_a^2+m_b^2\right)}\log(m_c^2/m_t^2) \\
   &+& \frac{m_a^2 \left(m_b^4+m_W^4\right)+2 m_b^2
   \left(m_b^2+m_W^2\right) \left(m_t^2+2
   m_W^2\right)}{8 m_a^2 m_b^3 m_W}\log(m_c^2/m_W^2) \\
   &+& \frac{m_s^4-m_s^2 \left(2
   m_t^2+m_W^2\right)+m_t^4+m_t^2 m_W^2-2
   m_W^4}{4 m_s m_W m_a^2}\log(m_t^2/m_W^2)+\frac{m_b}{4 m_W}\left(\frac{1}{\bar{\varepsilon}}+\log(\mu^2/m_c^2)\right)\\
   &+& \frac{m_b \left(m_t^4+m_t^2 m_W^2-2
   m_W^4\right)}{2 m_a^2 m_W} C_0(0, 0, m_a^2, 0, m_W, m_t) \\
   & + & \frac{m_s \left(m_s^2-m_t^2-2 m_W^2\right)}{2
   m_W m_a^2} B(m_s^2, m_t, m_W)
\end{eqnarray*}

 The two- and three-point functions are the following
 \begin{eqnarray*}
     B(M^2,m_1,m_2) &=& \frac{\lambda ^{1/2}\left(M^2,m_1^2,m_2^2\right)}{M^2} \log
   \left(\frac{\lambda
   ^{1/2}\left(M^2,m_1^2,m_2^2\right)-M^2+m_1^2+m_2^
   2}{2 m_1 m_2}\right) \\
   C_0(0, 0, m_a^2, 0, m_W, m_t) &=& \frac{\text{Li}_2\left(1-\frac{m_t^2}{m_W^2}\right)}{m_a^2
   }+\frac{\text{Li}_2\left(\frac{m_W^2}{m_a^2-m_t^2}+1\right)}{m_a^2}+\frac{\log
   ^2\left(-\frac{m_W^2}{m_a^2-m_t^2}\right)}{2
   m_a^2}
 \end{eqnarray*}
 where $\lambda(x,y,z)=x^2+y^2+z^2-2xy-2xz-2yz$ and $\text{Li}_2$ is the dilog function. These calculations were performed with the help of Package-X 2.0~\cite{Patel:2016fam}.

 \bibliography{myrefs}
\end{document}